\documentclass[12pt]{article}
\usepackage{graphicx}
\usepackage{epsfig}
\def\be{\begin{equation}}
\def\ee{\end{equation}}
\def\ba{\begin{array}}
\def\ea{\end{array}}
\def\bea{\begin{eqnarray}}
\def\eea{\end{eqnarray}}

\begin{document}

\title{New insight on pseudospin doublets in nuclei} 
\author{ 
B.  Desplanques$^{1}$\thanks{{\it E-mail address:}  desplanq@lpsc.in2p3.fr},
S. Marcos$^{2}$ \\ 
$^{1}$ LPSC, Universit\'e Joseph Fourier Grenoble 1, CNRS/IN2P3, INPG,\\
  F-38026 Grenoble Cedex, France \\ 
$^{2}$ Departamento de Fisica Moderna, Universidad de Cantabria, \\
E-39005 Santander, Spain  }

\sloppy

\maketitle

\begin{abstract}
\small{
The relevance of the pseudospin symmetry in nuclei is considered. 
New insight is obtained from looking at the continuous transition 
from a model satisfying the spin symmetry to another one satisfying 
the pseudospin symmetry. This study suggests that there are models 
allowing no missing single-particle states in this transition, 
contrary to what is usually advocated. It rather points out to an
association of pseudospin partners different from the one usually assumed, 
together with a strong violation of the corresponding symmetry. 
A comparison with results obtained from some relativistic approaches is made.}
\end{abstract} 
and strength functions \\
\noindent


\section{Introduction}
The concept of pseudospin symmetry (PSS) in the description of single-particle 
nuclear levels has  been introduced 40 years ago \cite{Arim69,Hech69}. 
It was motivated by the observation that levels within the same major shell
such as $2s_{1/2}$ and $1d_{3/2}$ ($N=2$), $2p_{3/2}$ and $1f_{5/2}$ ($N=3$), 
{\it etc}, are rather close to each other and, in any case,
closer than usual spin-orbit partners. Thus, the pseudospin symmetry 
could be more accurately satisfied than the spin symmetry (SS). 
Approximate pseudospin symmetry has been also found and analyzed in deformed
nuclei \cite{Bohr82}-\cite{Cas92} before a general explanation were suggested and
lengthily accepted.

Some theoretical explanation supporting the existence of pseudospin doublets 
was given by Ginocchio \cite{Gino97}, 
within the phenomenological relativistic framework originally proposed 
by Walecka for nuclear matter \cite{Wal74} and used later on for studying 
properties of finite nuclei  \cite{Bou84,Ser86}. Looking at the small components 
of the Dirac spinors \cite{Dir28}, he noticed that the equation allowing one 
to determine them  leads to degenerate pseudospin (PS) doublets 
in the limit where the sum of  the scalar ($\Sigma_S$) and vector  ($\Sigma_0$) 
components  of the nucleon self-energies vanishes. 
In this limit, the radial parts of the small components are identical. 
This result suggested him to propose that the approximate pseudospin symmetry 
is due to the fact that the sum of  the scalar and vector  self-energies, 
$\Sigma_S$ and $\Sigma_0$, cancels for a large part \cite{Gino98}-\cite{Gino05}. 
The attractive explanation of the pseudospin symmetry has motivated 
a considerable number of studies by other authors both to extend its result 
\cite{Gam98}-\cite{Hui06}
and to get a better understanding of its limitations 
\cite{Mar00}-\cite{Lop05b}.
Among these last ones, it was noticed that the quasi cancellation 
of the  self-energies in the sum $\Sigma_S+\Sigma_0$ 
was associated with an energy denominator which, actually, does cancel 
at the nuclear surface \cite{Mar00}-\cite{Mar07}, \cite{Alb02}.
This indicates that some care is required in concluding from the cancellation
of the $\Sigma_S$ and $\Sigma_0$ potentials alone \cite{Mar08}. 
Thus,  the expected similarity of the radial part of the small components 
of the Dirac spinors was not strongly supported by the detailed analysis 
of various solutions   \cite{Mar04,Mar07}. It even turned out that 
the similarity of the radial part of the large components  
for spin-orbit partners was better satisfied while the conditions 
for the energy levels required by the symmetry 
in this case are apparently worse \cite{Mar04,Mar07}. 
Last, but not least, the pseudospin symmetry supposes the existence 
of single-particle states that are not observed \cite{Lev01},
{\it i.e.} the PS partners of the so-called intruder states 
(states with $n=1$ and $j=l+1$).
For clarifying some of the above points, a model where the above mentioned 
energy denominator does not cancel would be useful. In this case, however, 
any realistic nuclear model would not admit bound states.

In this work, we mainly concentrate on the possible missing states. 
For our purpose, we first work in the non-relativistic case and consider 
two interactions that evidence the spin and pseudospin symmetry. 
For simplicity, we assume that they are related to each other 
by a unitary transformation that commutes with the kinetic-energy operator. 
To study specifically the missing-state problem, we also introduce 
an interaction that is a superposition of the two above ones 
with  relative weights, $1-x$ and $x$. By varying the $x$ parameter 
from $0$ to $1$, one should see how energy levels in the spin-symmetry case 
are related (or unrelated) to those in the pseudospin-symmetry one 
and {\it vice versa}. Some extension to a relativistic approach is made 
with the idea of an application to clarifying Chen {\it et al.} 
results \cite{Chen02}, which evidence a spectrum very similar to ours 
in the nonrelativistic approach. Tentatively, we also considered 
the case of finite potentials instead of harmonic-oscillator-type ones 
used in  the above work.

The plan of the paper is as followed. The second section is devoted to the
general aspects of the non-relativistic interaction we are using for our study. 
Results for spectra obtained from two different interaction models 
are presented in the third section  together with some discussion.  
Complementary information from examining wave functions is given 
in  the fourth section. The fifth section is devoted to a comparison 
with results from  relativistic approaches. Conclusions are given 
in the sixth section. Two appendices contain some technical details.

\section{Interaction model}
The interactions with spin and pseudospin symmetry we start with 
may be written as:
\begin{eqnarray}
&&V_0=V_0(r)\, ,
\nonumber \\
&& \tilde{V}_1=
 \vec{\sigma} \!\cdot\! \hat{p} \, V_1(r)\, \vec{\sigma} \!\cdot\! \hat{p}\, ,
\end{eqnarray}
where $V_0(r)$ and $V_1(r)$ are spin independent and only depend 
on the $r$ variable (local potentials). The operator 
$ \vec{\sigma} \!\cdot\! \hat{p}$ appearing in  $\tilde{V}_1$ is an unitary one.
It changes the states corresponding to the pseudospin-symmetry space 
(represented with the symbol $\tilde{}\,$)  to those
corresponding to the spin-symmetry space \cite{Blo95}. 
In this operation, the total angular momentum, $j$, is preserved 
while the orbital angular momentum is modified by one unit,
implying a change of parity. Moreover, the operator makes the interaction 
$\tilde{V}_1$ non local.
The corresponding Hamiltonian may therefore be written as:
\begin{eqnarray}
\tilde{H}_1=\frac{p^2}{2M}+\tilde{V}_1=
\vec{\sigma} \!\cdot\! \hat{p} \, 
\Big( \frac{p^2}{2M}+V_1(r) \Big) \, \vec{\sigma} \!\cdot\! \hat{p} \,,
 \end{eqnarray}
indicating that the spectra corresponding to the interactions 
$ \vec{\sigma} \!\cdot\! \hat{p} \, V_1(r)\, \vec{\sigma} \!\cdot\! \hat{p}$ 
and $V_1(r)$ will be the same, apart from the spin-orbital angular momentum 
assignments of course. Wave functions for the Hamiltonian $\tilde{H}_1$ can be
obtained from those for the Hamiltonian $H_1= \frac{p^2}{2M}+V_1(r)$ 
by applying to these ones the operator $ \vec{\sigma} \cdot \hat{p}$. 
Apart from the angular-momentum factor, the solutions of $H_1$ 
for spin-orbit partners are the same in both configuration and momentum spaces. 
As the operator  $ \vec{\sigma} \!\cdot\! \hat{p}$ is local in momentum space, 
this equality  will turn into an equality for momentum-space solutions  
of $\tilde{H}_1$  corresponding to pseudospin partners. 

The total interaction that is useful for our purpose of looking at the
transition of the spin-symmetry case to the pseudospin-symmetry one 
may be written as: 
\begin{eqnarray}
V(r)&=& (1-x) \,  V_0 + x\, \tilde{V}_1
\nonumber \\
&=& (1-x) \, V_0(r)
+x\, \vec{\sigma} \!\cdot\! \hat{p} \, V_1(r)\, \vec{\sigma} \!\cdot\! \hat{p}\, ,
\label{eq:vnr}
\end{eqnarray}
where $x$ is supposed to vary from  $0$ to $1$. Actually, 
as the interaction $\tilde{V}_1$ is spin dependent,
we can recover a part of the usual spin-orbit splitting by taking 
slightly negative values of $x$.
\begin{table}
\begin{center}
\begin{tabular}{lccc}
\hline  \vspace*{-3mm}\\ 
& spin symmetry  & &  pseudospin symmetry \\  [1.ex] 
\hline  \vspace*{-3mm}\\
& $2p_{1/2},\;2p_{3/2},\;1f_{5/2},\;1f_{7/2}$ & $\leftrightarrow$ &
$2\tilde{s}_{1/2},\;2\tilde{d}_{3/2},\;\;\;1\tilde{d}_{5/2},\;1\tilde{g}_{7/2}$ 
\\  [0.5ex]
& $2s_{1/2},\;1d_{3/2},\;1d_{5/2}$ & $\leftrightarrow$ & 
$2\tilde{p}_{1/2},\;\;\;1\tilde{p}_{3/2},\;1\tilde{f}_{5/2}$  \\  [0.5ex]
& $1p_{1/2},\;1p_{3/2}$ & $\leftrightarrow$ &
$1\tilde{s}_{1/2},\;1\tilde{d}_{3/2}$ \\  [0.5ex]
& $1s_{1/2}$ & $\leftrightarrow$ & $1\tilde{p}_{1/2}$\\  [1.ex]
\hline
\end{tabular}
\end{center}
\caption{Representation of spectra in the spin-symmetry and 
pseudospin-symmetry limit (l.h.s. and r.h.s. respectively): 
states on the left in each row correspond to standard major shells. 
For the interaction assumption $V_1(r)=V_0(r)$ considered in this work, 
states on the r.h.s. in a given row have the same energy 
as those on the l.h.s. in the same row and the same total angular momentum 
(but a different parity). We anticipated that the quantum number $n$ is
preserved in making this correspondence.
\label{table:spectrum} }
\end{table}

As mentioned in the introduction, we consider the particular case where 
$V_1(r)=V_0(r)$. This does not diminish the relevance of  results to be obtained
here, making them more striking instead. Without performing any calculation, 
one can see that the states for the spin and pseudospin-symmetry cases, 
shown respectively in the l.h.s. and r.h.s. of  Table \ref{table:spectrum}, 
are in a one-to-one correspondence. The $n$ quantum number represents the order of
the states and, apart from a possible conventional shift by one unit,  
it is not, necessarily, identical to the node number of the radial wave function. 
This relationship, which is often assumed implicitly in the literature,
is specific of a local potential. 
The question of interest is whether there is a global continuity 
between states on the l.h.s. of the table and those on the r.h.s. 
when the interaction varies from the spin-symmetry limit
($x=0$) to the pseudospin-symmetry one ($x=1$). 
It is reminded that the current understanding of the pseudospin symmetry 
supposes that the  pseudospin partners of the states on the very right 
in each row of the left part of  Table \ref{table:spectrum}  are absent, 
due to specific conditions of realistic nuclear potentials \cite{Lev01}. 
The argument relies on a mathematical analysis of the nodes of wave functions. 
The authors do not examine in what measure this absence could have
a relation with the differences between realistic models and those satisfying
exact PSS. As we shall see in this work, as one approaches the PSS limit, 
a new type of nodes could appear, which may affect their conclusion.

It remains to specify the choice of the potential $V_0(r)(=V_1(r))$. 
We consider two choices inspired by the harmonic-oscillator 
and the Woods-Saxon potentials. The comparison of results for these two models
evidences a little but important detail that could be relevant for the
interpretation of pseudospin doublets. The absence of singular behavior for
these potentials lets one to expect a continuous transition between states 
on the l.h.s. and r.h.s. of Table  \ref{table:spectrum}.

\section{Results and discussion for the spectra}
\begin{figure}[htb]
\mbox{ \epsfig{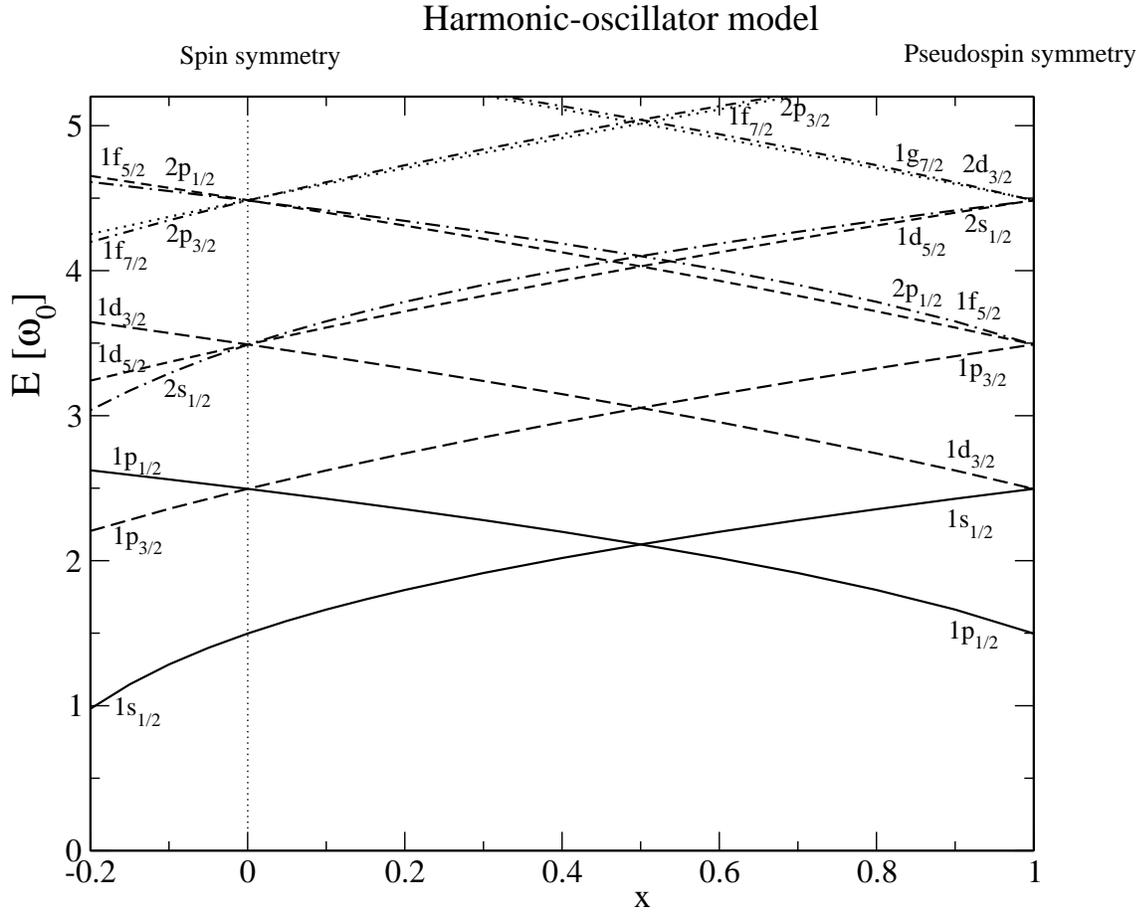}}
\caption{Single-particle spectrum evidencing the spin- and pseudospin-symmetry
cases and the continuous transition between them. The interaction is based 
on a harmonic-oscillator one:
$ V=\frac{1}{2} M \omega_0^2 \;\Big((1\!-\!x)\;r^2 
+ x \;\vec{\sigma} \!\cdot\! \hat{p}\; r^2 \;\vec{\sigma} \!\cdot\! \hat{p} \Big)$.
The observed order of the spin-orbit partners is recovered for negative
values of $x$. The figure includes on the l.h.s. two ``standard pseudospin 
doublets'', $1d_{3/2}$ and $2s_{1/2}$ on the one hand, $1f_{5/2}$ 
and $2 p_{3/2}$ on the other hand.
\label{fig_ho}}
\end{figure}  
We, successively, present in this section level spectra for the harmonic-oscillator 
(HO) and the Woods-Saxon (WS) models. Parameters are those 
appropriate to the description of a nucleus such as $^{40}Ca$. 
This is followed by some comments about the spectra.
\subsection{Harmonic-oscillator model}
For the potential $V(r)$ inspired from  the harmonic-oscillator form, 
the two terms $V_0(r)$ and $V_1(r)$ entering its definition assume 
the standard form:
\begin{equation}
V_0(r)=V_1(r)= \frac{1}{2} M \omega_0^2 \, r^2 \, .
\end{equation}
As results for the energy levels only depend on the parameter $\omega_0$, 
it is sufficient to present them  in units of this parameter. 
They are shown in Fig. \ref{fig_ho} for $x$ varying from -0.2 to 1. 
Actually, calculations were done with a Gaussian potential deep enough 
to be approximated by the HO potential. The reason is that 
this is more convenient for our study, which is made in the momentum space  
due to the non-locality of the interaction model we are using when $x \neq 0$. 
Moreover, the Fourier transform of the Gaussian potential can easily be
performed. As can be observed from the figure,
there is almost no departure with the results of the harmonic-oscillator model
when a comparison is possible (at $x=0$). While the spectra can be essentially
given in terms of  $\omega_0$, the value taken by this parameter is 
nevertheless  relevant in obtaining wave functions that could be compared 
to those of the WS model. We use $\omega_0=7$ MeV so that the spectra 
of the HO and WS models roughly overlap.

Examination of the spectrum around $x=-0.2$ shows it {\it a  priori} 
contains the two standard pseudospin doublets, $1d_{3/2},\; 2s_{1/2}$ 
and $1f_{5/2},\;2p_{3/2}$, though the corresponding states are slightly apart 
in each case. The comparison with the spectrum at $x=1$, which evidences an
exact pseudospin symmetry, suggests a different pattern however.
The states $1d_{3/2}$ and $ 2s_{1/2}$ (or $1f_{5/2}$ and $2p_{3/2}$) 
tend to separate in this limit while the states 
$1d_{3/2}$ and $ 1s_{1/2}$ (or $1f_{5/2}$ and $1p_{3/2}$) get closer, 
supporting that these ones should be associated in the same pseudospin doublet.
We notice that the assignment
of the $n$ quantum number to the different states is made by continuity 
with results   at $x=0$. In this case, where the potential is local, 
there is a one-to-one correspondence between this number and the number of
nodes in the radial wave function. This correspondence is lost as soon as the
interaction is partly non-local. Examination of radial wave functions like the
$1s_{1/2}$ for $x=1$ thus evidences a node. This one occurs however in the tail
of the wave function, at a distance where the contribution 
to the normalization is almost saturated (within a few percent's). It differs
from standard nodes that allow one to determine contributions to the
normalization from smaller and larger $r$ of comparable sizes. 
Only these ones could have some relation to the quantum number $n$. 
We also notice that the pseudospin-symmetry spectrum obtained for $x=1$ 
shows the same degeneracy pattern as the one obtained by Chen  {\it et al.} 
in a relativistic approach with a harmonic-oscillator-type interaction
\cite{Chen02}. Whether there is a relationship between the two cases 
will be discussed in Sect. 5.
\subsection{Woods-Saxon model}
\begin{figure}[htb]
\mbox{ \epsfig{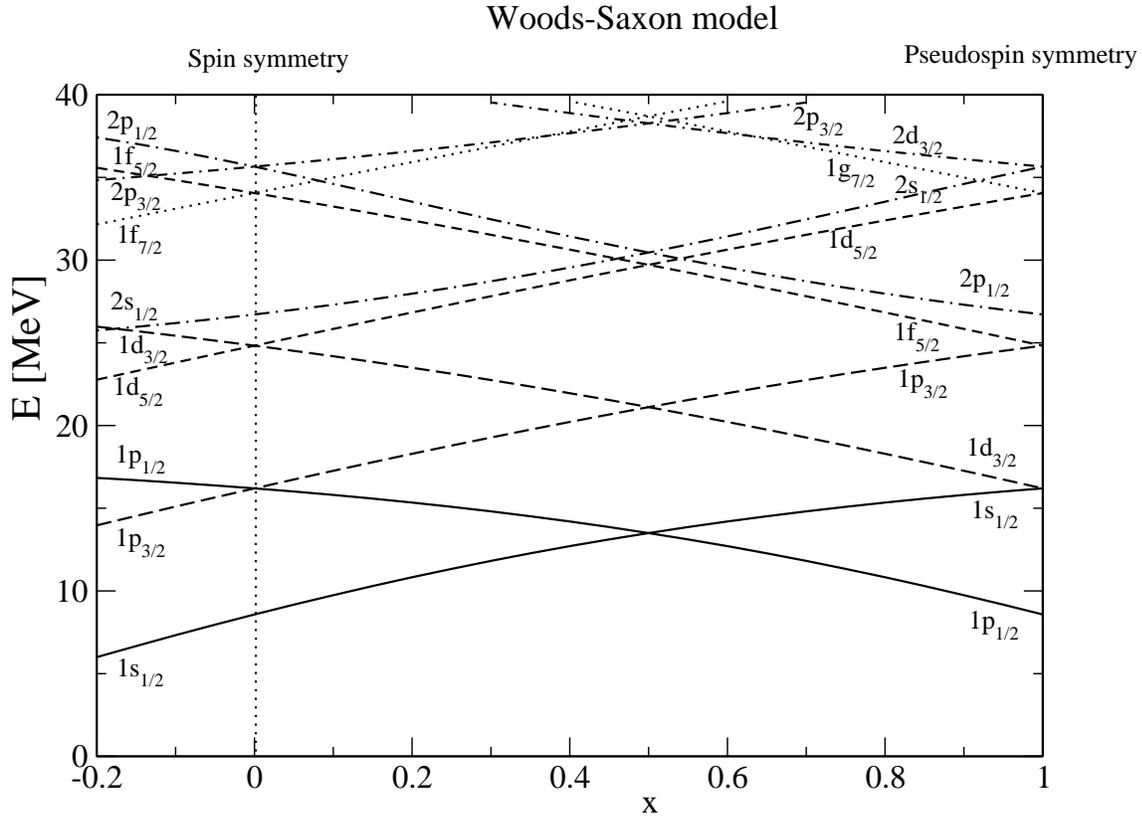}}
\caption{Same as in Fig. \ref{fig_ho}, but for the Woods-Saxon model 
\label{fig_ws}}
\end{figure}  
For the potential $V(r)$ inspired from  the Woods-Saxon form, 
the two terms $V_0(r)$ and $V_1(r)$ entering its definition assume 
the standard form:
\begin{equation}
V_0(r)=V_1(r)= -\frac{V_0}{1+{\rm exp}((r-r_0)/a))}  \, .
\end{equation}
\begin{figure}[htb]
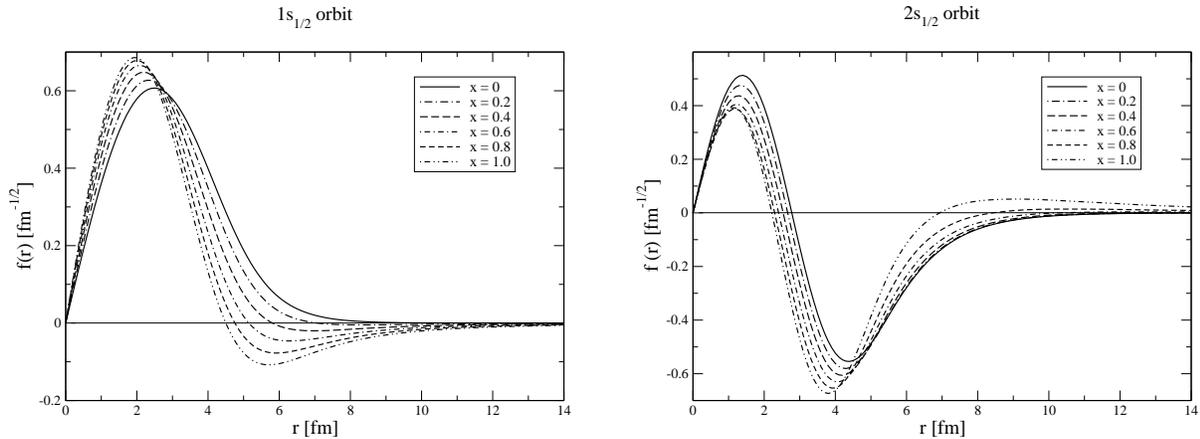

\mbox{\epsfig{ file=fnsgs0u.eps, width=7.5cm}\hspace*{0.7cm}
\epsfig{ file=fnsex0u.eps, width=7.5cm}}
\caption{Appearance of an extra node when going from the spin- 
to the pseudospin-symmetry limit for orbits $1s_{1/2}$ and $2s_{1/2}$ 
(left and right panels respectively) with the Woods-Saxon model. It is noticed that the
extra node appears in the tail of the wave function and cannot be confused with
the usual nodes that appear at smaller distances, as the comparison of the two
panels can show. The wave functions are normalized as $\int dr\, f^2(r)=1$.
\vspace*{1.0cm} \label{node1}}
\end{figure}  
\begin{figure}[htb]
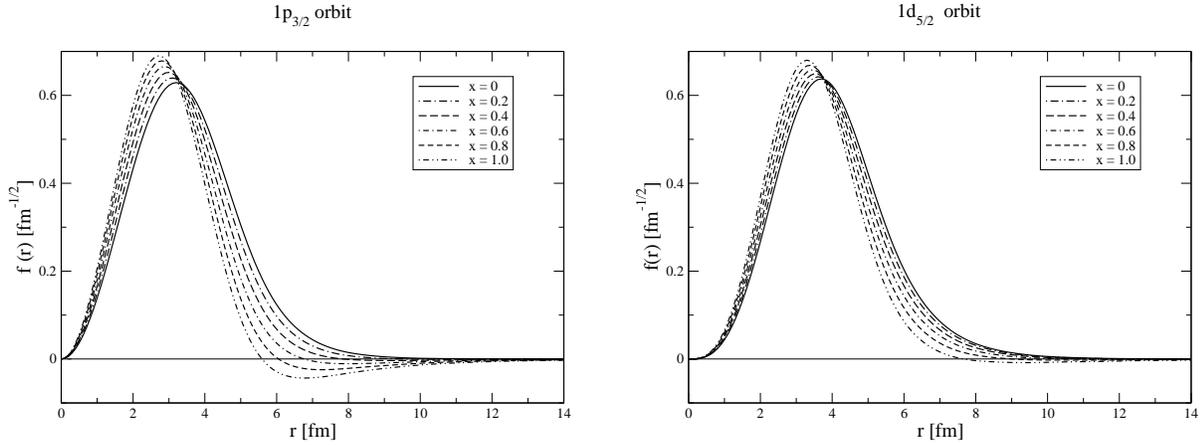

\mbox{\epsfig{ file=fnsgs1u.eps, width=7.5cm}\hspace*{0.7cm}
\epsfig{ file=fnsgs2u.eps, width=7.5cm}}
\caption{Same as for Fig. \ref{node1} but for orbits $1p_{3/2}$ and $1d_{5/2}$ 
(left and right panels respectively).
\label{node2}}
\end{figure}  
For numerical calculations, we use $V_0=40$ MeV, $r_0=5$ fm, $a=0.65$ fm. 
Results for the spectrum are shown in Fig. \ref{fig_ws}.
Examination of this figure confirms most of the features observed 
for the HO model. The energies of the two
members of the pseudospin symmetry doublets,  $1d_{3/2},\; 2s_{1/2}$ 
and $1f_{5/2},\;2p_{3/2}$, are closer to each other however. 
As for the HO model, the quantum number, $n$, has been assigned by continuity 
with the spin-symmetry case. This quantum number is not necessarily related 
to the number of nodes as often assumed implicitly. Thus, orbits with
$j=l+\frac{1}{2}$ tend to get an extra node when going gradually from the spin- 
to the pseudospin-symmetry limit while orbits with $j=l-\frac{1}{2}$ don't. 
The extra node occurs at the limit or outside the range 
of the potential $V_1(r)$. This is illustrated in Fig. \ref{node1} for the 
orbits  $1s_{1/2}$ and $2s_{1/2}$  and in Fig. \ref{node2} for the orbits 
$1p_{3/2}$ and $1d_{5/2}$. The contribution to the norm beyond the extra node 
amounts to 3\% in the largest case and tend to decrease either with going to the
spin-symmetry limit or with increasing the orbital angular momentum. 
In comparison, the contribution to the norm for a usual node
amounts to several 10\% (see the two panels of Fig. \ref{node1} 
for a $s_{1/2}$ orbit).
\subsection{Comments}
Examination of the spectra for two interaction models shows that there is
essentially a one-to-one global correspondence between states with the spin
symmetry ($x=0$) and those with the pseudospin symmetry ($x=1$). 
There is no indication for missing pseudospin partners as often advocated. 
Instead, all states are there but in a different order. The order for the 
pseudospin-symmetry states may be a surprise but this is a consequence of the
non locality of the corresponding interaction used here.

The interaction model we used does not contain explicit spin-orbit interaction 
but it is found that the nonlocal term proportional to $V_1$ provides a similar
effect which, for $x=-0.2$, has the appropriate size. This is not unexpected as
there is some similarity of this term with the one that produces the effect in
relativistic mean-field approaches where it is proportional 
to $\vec{\sigma} \!\cdot\! \vec{p} \; V_1(r)\, \vec{\sigma} \!\cdot\!\vec{p} $.
We also notice that the actual description of nuclei ($x=-0.2$) is closer 
to the PS limit ($x=0$) than to the PSS limit ($x=1.0$) and, moreover, 
requires an opposite-sign value of $x$.

Our last comment is motivated by the comparison of the two models concerning the
standard pseudospin doublets. While results for the WS model at  $x=-0.2$
supports such doublets, this is not the case for the HO model. It is known from
the Nilsson model that a term proportional to $-\kappa \;\mu \;\vec{l}\,^2$ could be
necessary to improve the single-particle spectrum, lowering, for instance, 
the energy of the $1d_{3/2},\; 1d_{5/2}$ states with respect 
to the $2s_{1/2}$ one. This feature shows that the form of the potential is
important in getting the standard pseudospin doublets. 
This contrasts  with the explanation of these doublets 
in the relativistic mean-field approach \cite{Gino97}, 
which assumes that the approximate PS symmetry observed in this approach 
is, essentially, a direct consequence of the small size of
$\Sigma_S+\Sigma_0$ (in comparison with the magnitude of $\Sigma_S-\Sigma_0$ 
\cite{Gino99}), independently of its form.

\section{Results and discussion for the wave functions}
In discussing  pseudospin doublets, 
examination of the corresponding wave functions is thought to provide an
important information on the relevance of the possible underlying symmetry. 
In this section, we present results based on wave functions 
for the standard pseudospin-symmetry partners as well as those resulting 
from our interpretation of the doublets. This is done in both momentum and
configuration space, for the two interaction models we considered 
and for the two doublets that their spectra shown 
in Figs.  \ref{fig_ho} and  \ref{fig_ws} exhibit.

\begin{figure}[htb]
\mbox{\epsfig{ file=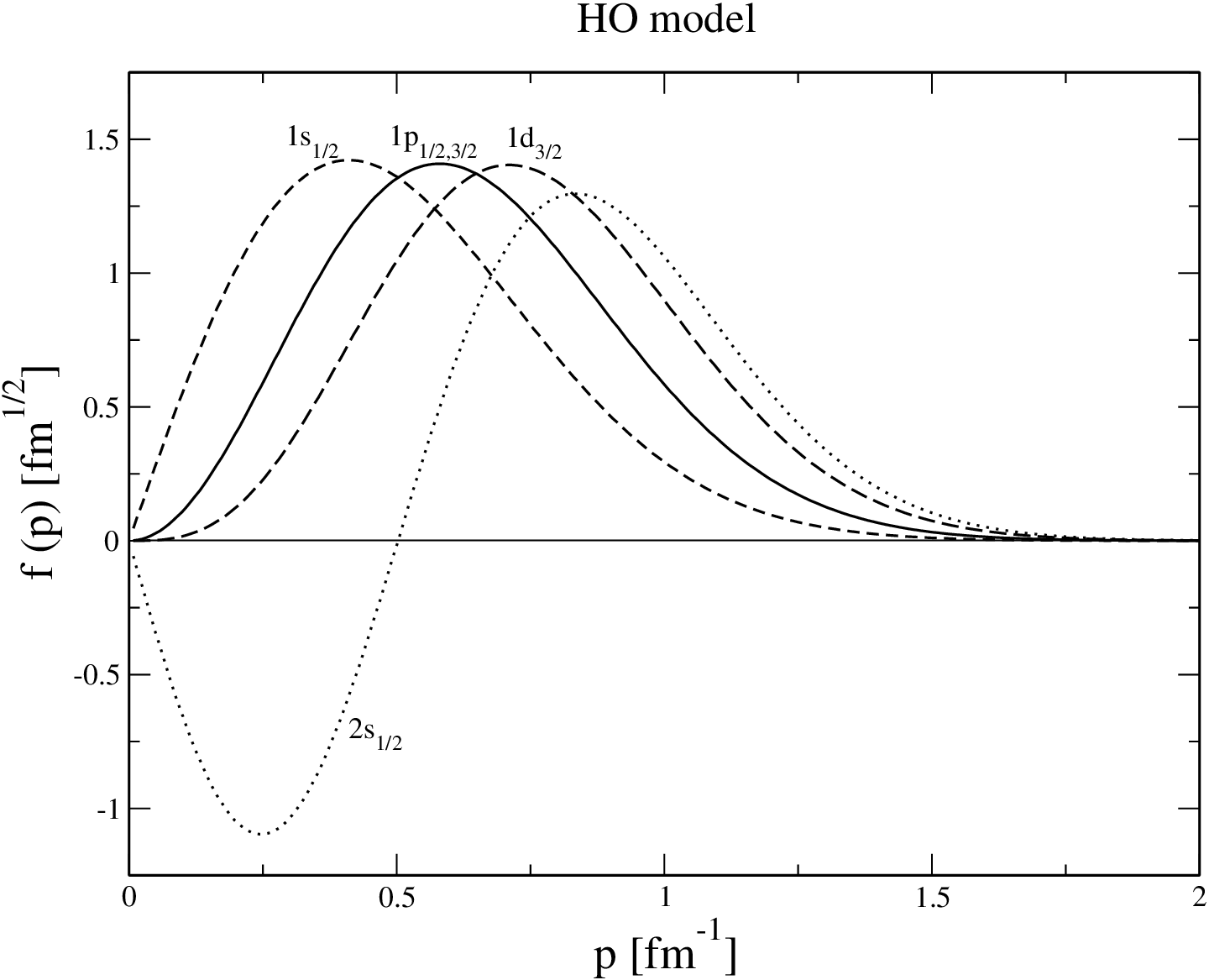, width=7.5cm}\hspace*{0.5cm}
\epsfig{ file=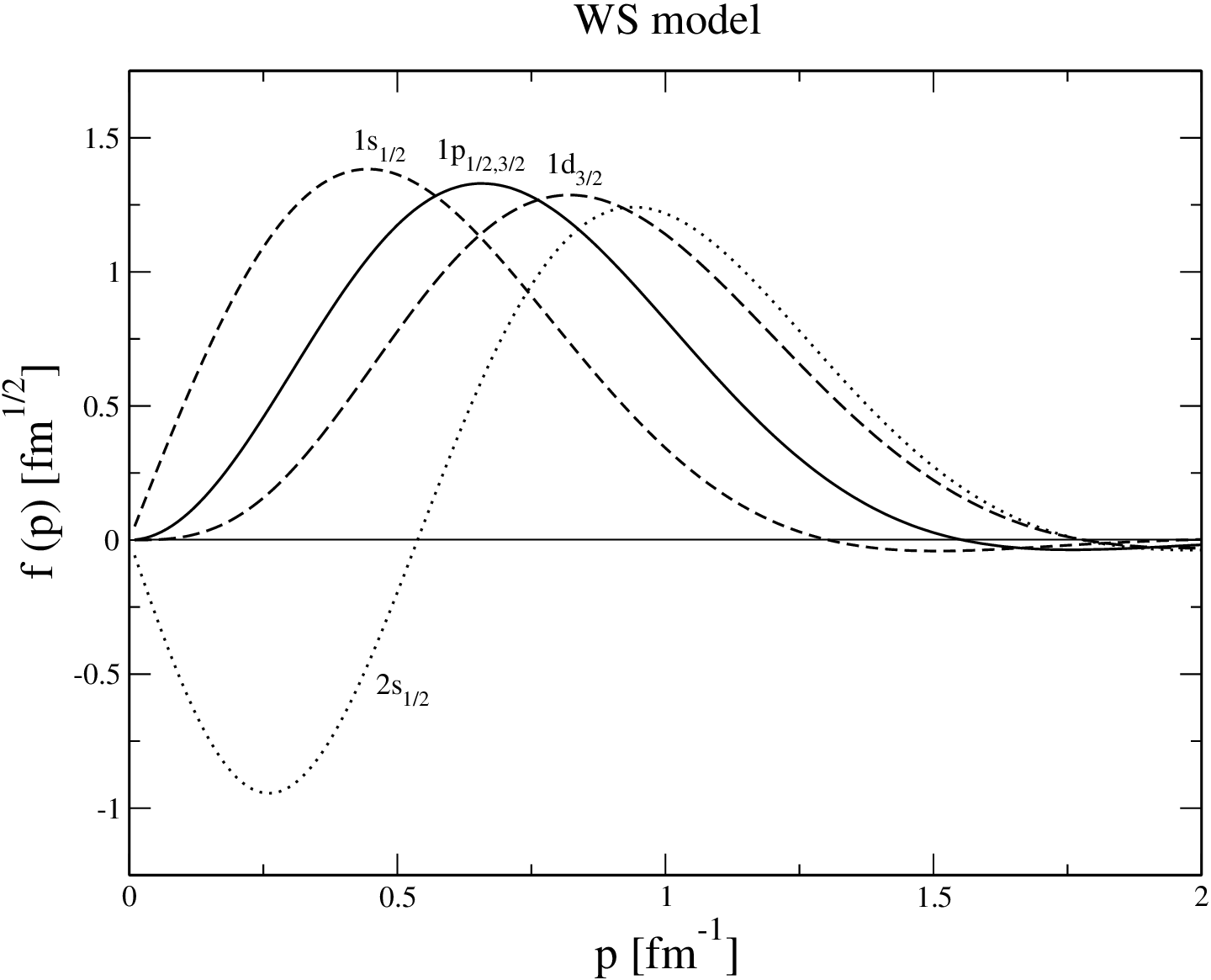, width=7.5cm}}
\caption{Comparison of momentum-space wave functions, $f(p)$, for the lowest
pseudospin doublet with the HO and WS models (left and right panels
respectively): small- and large-dash lines for the states $1s_{1/2},\;
1d_{3/2}$, continuous line for the limit expected in the pseudospin-symmetry
limit and dotted line for the $2s_{1/2}$ state. All wave functions shown 
in the figure are calculated at $x=0$.
\label{psp1}}
\end{figure}  
\begin{figure}[htb]
\mbox{\epsfig{ file=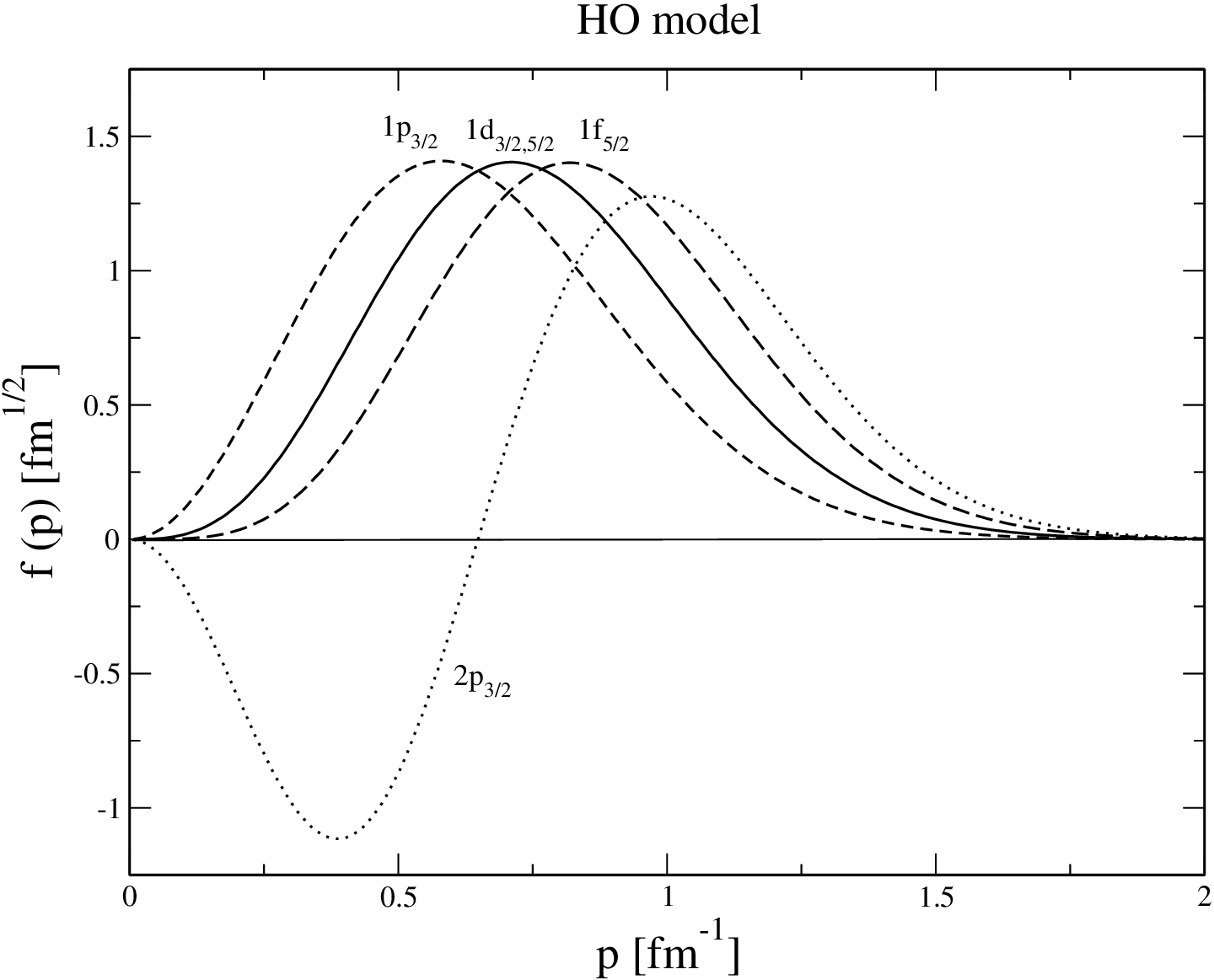, width=7.5cm}\hspace*{0.5cm}
\epsfig{ file=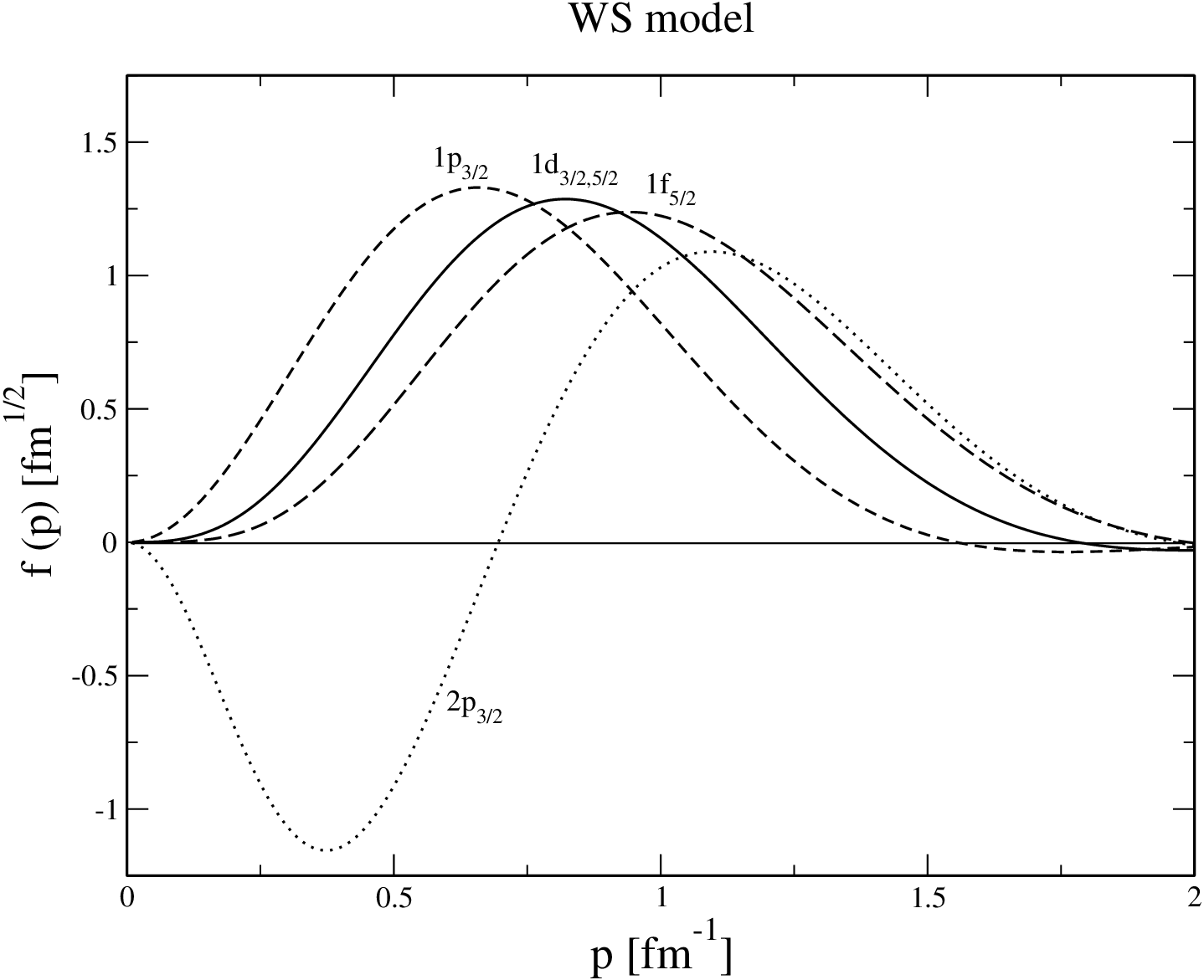, width=7.5cm}}
\caption{Comparison of momentum-space wave functions, $f(p)$, for the second lowest
pseudospin doublet with the HO and WS models (left and right panels
respectively): small- and large-dash lines for the states $1p_{3/2},\;
1f_{5/2}$, continuous line for the curve expected in the pseudospin-symmetry
limit and dotted line for the $2p_{3/2}$ state. All wave functions shown 
in the figure are calculated at $x=0$.
\label{psp2}}
\end{figure}  
\subsection{Results in momentum space}
The transformation of the usual spin space to the pseudospin one, 
given by the operator $\vec{\sigma}  \cdot  \hat{p}$, 
is local in the momentum space,  where its effect can  therefore be
more easily calculated. Thus, the spin-orbital angular momentum part of wave functions 
for the states $d_{3/2},\; s_{1/2}$  (or $f_{5/2},\;p_{3/2}$) transform into 
that for the states $\tilde{p}_{3/2},\; \tilde{p}_{1/2}$  
(or $\tilde{d}_{5/2},\;\tilde{d}_{3/2}$),
leaving unaffected the dependence on the modulus of $\vec{p}$, which we denote
$f(p)$. In the pseudospin-symmetry limit, 
these functions for the states $\tilde{p}_{3/2},\; \tilde{p}_{1/2}$  
(or $\tilde{d}_{5/2},\;\tilde{d}_{3/2}$) are the same. Hence, in this limit,
the functions $f(p)$ for the states $d_{3/2},\; s_{1/2}$  
(or $f_{5/2},\;p_{3/2}$) should be equal. Moreover, the  functions 
for the states $\tilde{p}_{3/2},\; \tilde{p}_{1/2}$  
(or $\tilde{d}_{5/2},\;\tilde{d}_{3/2}$) can be calculated in this limit. 
For our interaction model, it is simply given by the  functions 
for the states $p_{3/2},\; p_{1/2}$  (or $d_{5/2},\;d_{3/2}$) 
in the spin-symmetry limit. We thus get relations like the following ones:
\begin{eqnarray}
f_{1d_{3/2}}(p)=f_{1s_{1/2}}(p)=f_{1p_{1/2,3/2}}(p)\,,
\nonumber \\
f_{1f_{5/2}}(p)=f_{1p_{3/2}}(p)=f_{1d_{3/2,5/2}}(p)\,.
\label{eq:relp}
\end{eqnarray}
One can therefore compare the $f(p)$ functions 
for the pseudospin doublet, according to our interpretation, 
$1d_{3/2},\; 1s_{1/2}$  (or $1f_{5/2},\;1p_{3/2}$), 
for the spin doublet $1p_{3/2},\; 1p_{1/2}$  (or $1d_{5/2},\;1d_{3/2}$), 
and  for the standard pseudospin doublet 
$1d_{3/2},\; 2s_{1/2}$  (or $1f_{5/2},\;2p_{3/2}$) (where we substituted
the state $2s_{1/2}$ (or $2p_{3/2}$) for the state $1s_{1/2}$ (or $1p_{3/2}$)). 
For the purpose of the
comparison, the functions for the pseudospin doublet   
$d_{3/2},\; s_{1/2}$  (or $f_{5/2},\;p_{3/2}$) are calculated 
at $x=0$. They fulfill the following normalization condition:
\begin{equation}
\int dp\; f^2(p) =1\,.
\label{eq:normp}
\end{equation}
The corresponding results are shown in Figs. \ref{psp1} and  \ref{psp2} 
for the lowest and next to the lowest pseudospin doublet respectively. 
In each case, results for the two interaction models are shown.

A quick look at the results with the two interaction models, 
as well as for the two pseudospin doublets, according to our definition, 
does not evidence much qualitative difference 
in the range which mainly contributes to the normalization.
As the pseudospin symmetry is not fulfilled at the point $x=0$ 
where the functions $p_{3/2},\; p_{1/2}$  (or $d_{5/2},\;d_{3/2}$) are
calculated, one expects that the various curves shown in the figures evidence
some discrepancy. With this respect, these discrepancies are not small. 
At low $p$, they essentially reflect the $p^{l+1}$ 
behavior expected for a local interaction model. At higher $p$, they rather
reflect the average momentum of the state, which increases with its energy  as
measured from the bottom of the potential well. It is nevertheless seen that
functions  $f(p)$ for the pseudospin doublets, as considered here, have a unique
sign in the range which mainly contributes to the norm. Moreover, they share
this feature with the function expected in the pseudospin symmetry limit. 
Instead, the functions for the standard pseudospin doublets evidence 
a different structure in the same range. One of the function 
has a given sign while the other one (with a different
$n$ number) evidences a change in sign. It thus turns out 
that our assignment for pseudospin doublets is more in agreement 
with the pseudospin-symmetry expectation than the standard assignment.
 
\subsection{Results in configuration space}
\begin{figure}[htb]
\mbox{\epsfig{ file=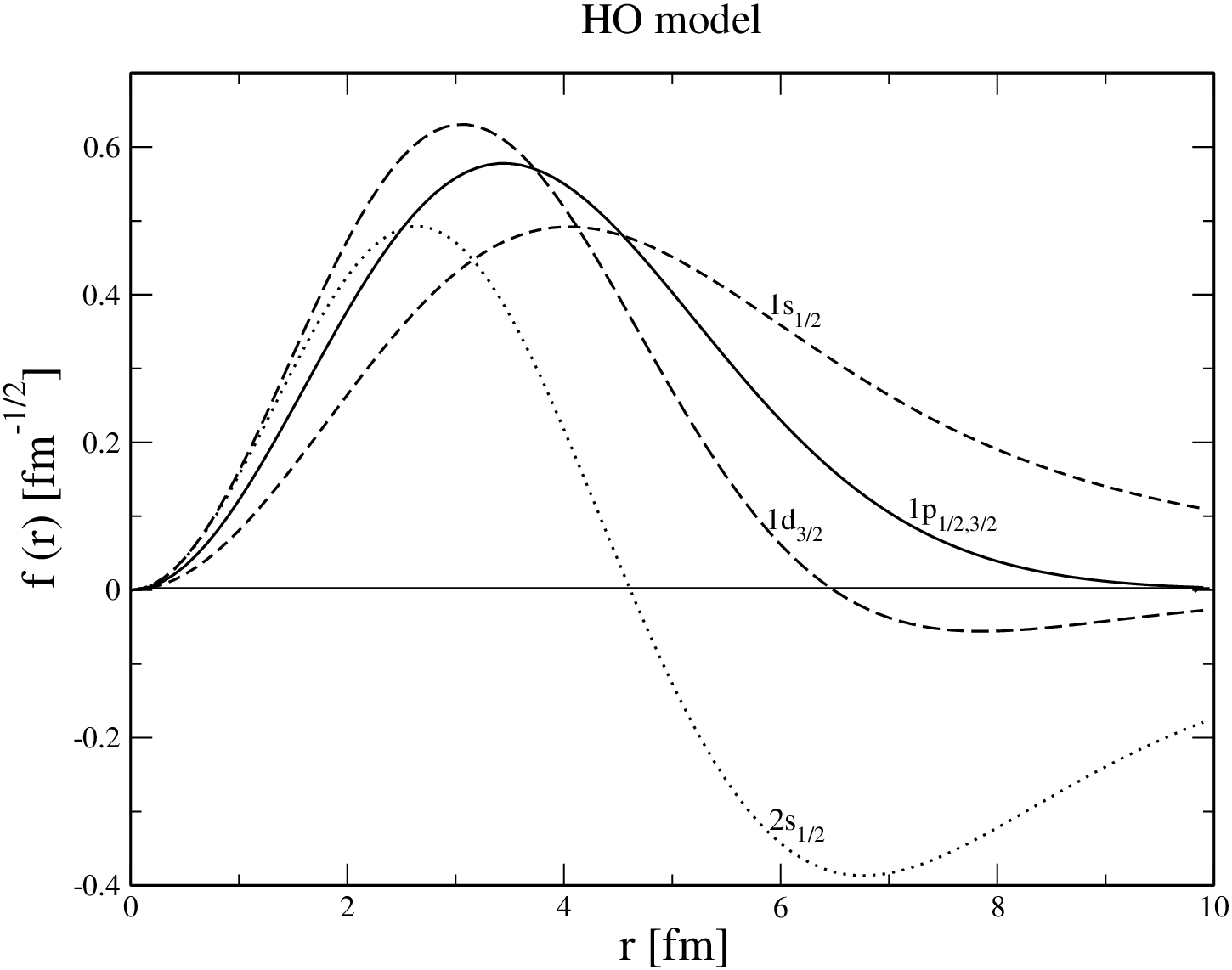, width=7.5cm}\hspace*{0.5cm}
\epsfig{ file=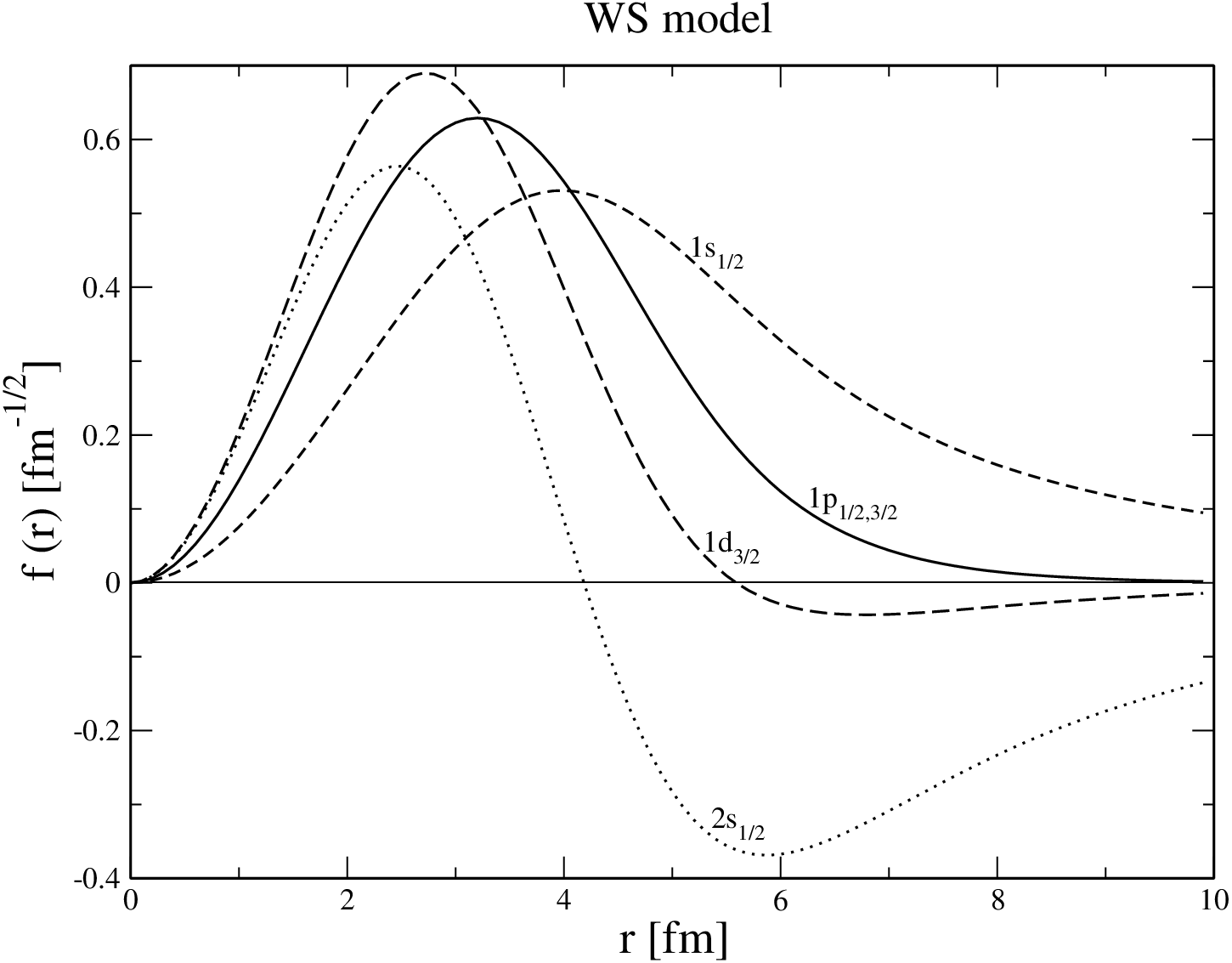, width=7.5cm}}
\caption{Comparison of configuration-space wave functions for the lowest
pseudospin doublet with the HO and WS models (left and right panels
respectively). See caption of Fig. \ref{psp1} for the definition 
of the different curves.  \label{psr1}}
\end{figure}  
\begin{figure}[htb]
\mbox{\epsfig{ file=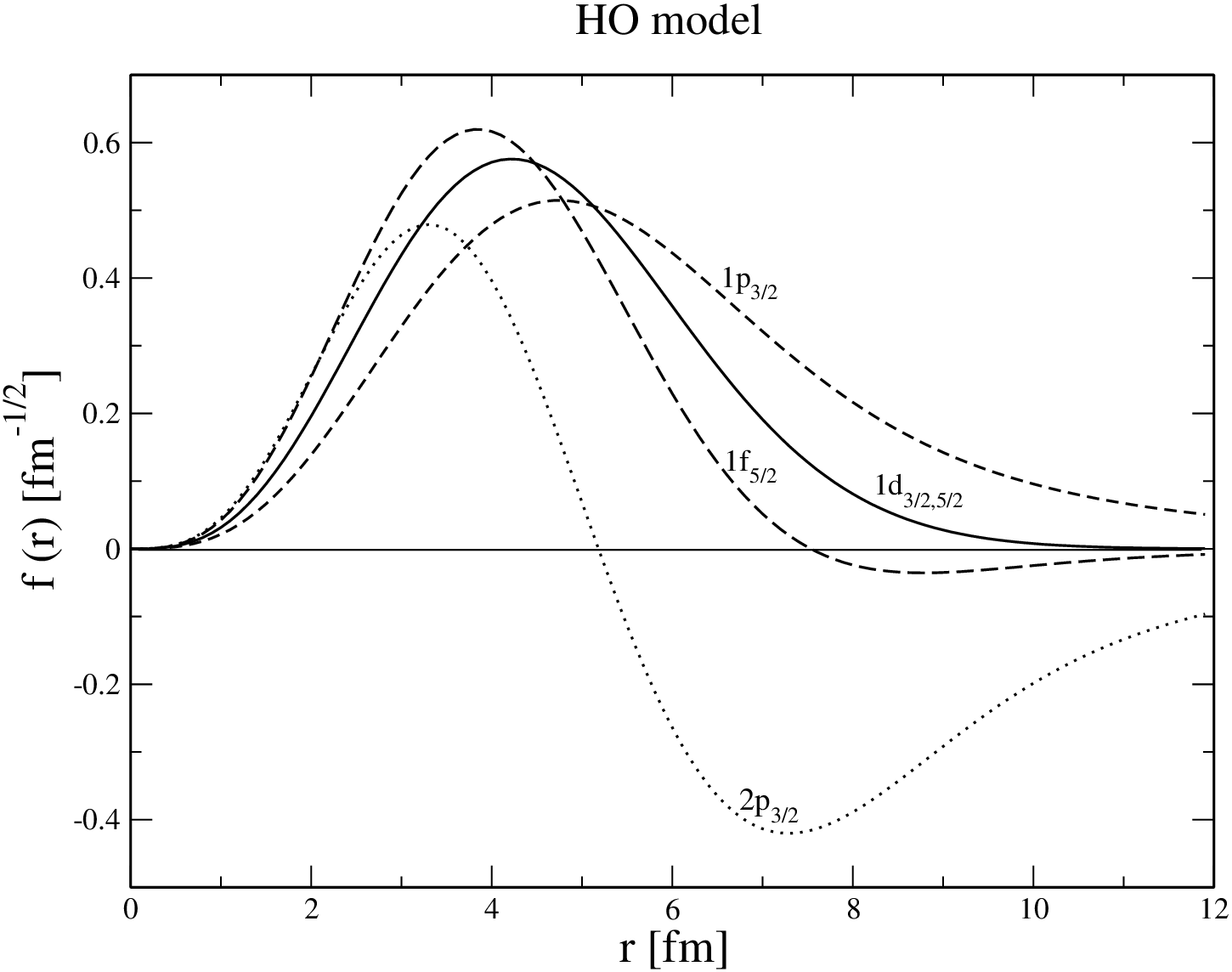, width=7.5cm}\hspace*{0.5cm}
\epsfig{ file=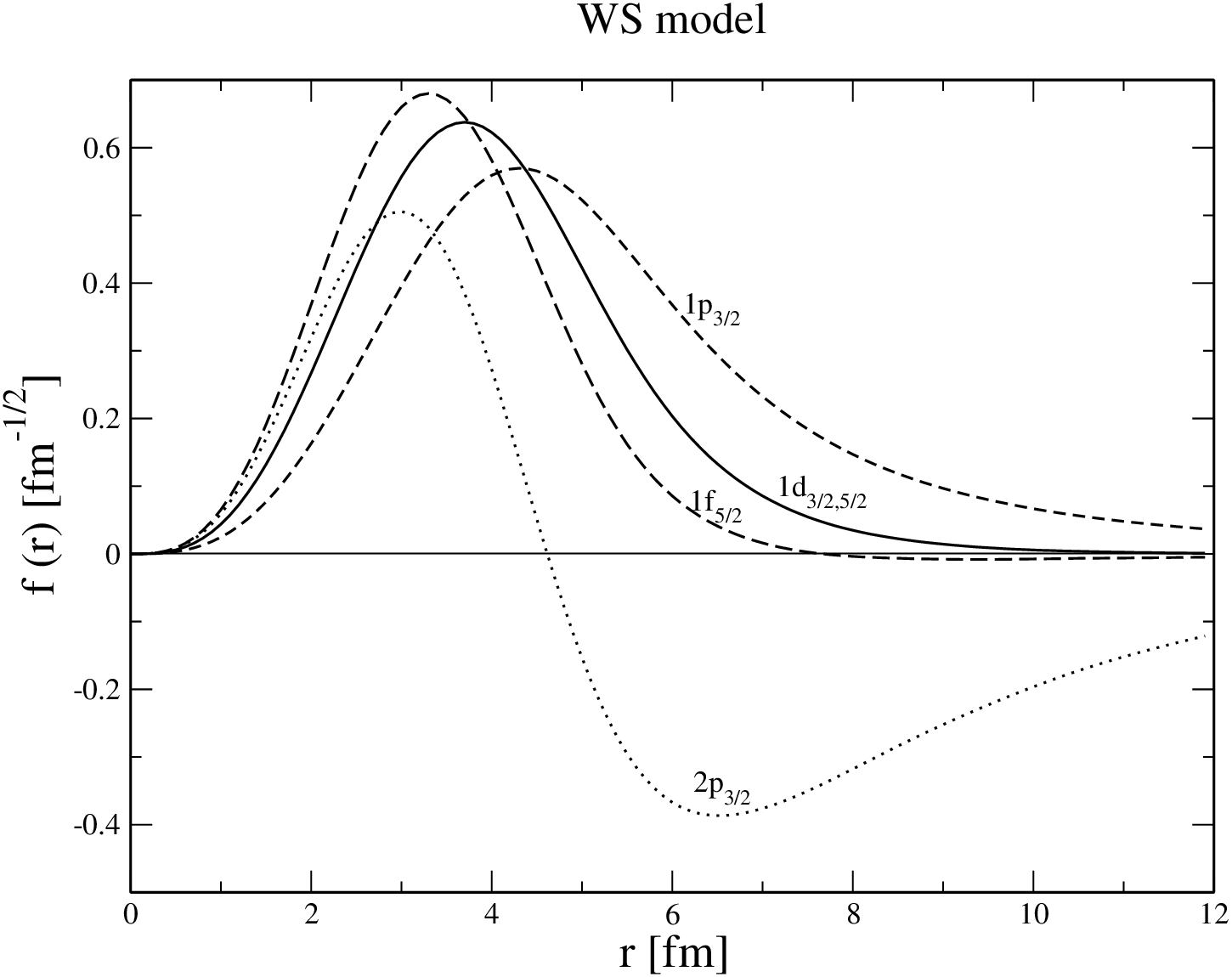, width=7.5cm}}
\caption{Comparison of configuration-space wave functions for the second lowest
pseudospin doublet with the HO and WS models (left and right panels
respectively). See caption of Fig. \ref{psp2} for the definition 
of the different curves.\label{psr2}}
\end{figure}  
While  the direct comparison of wave functions in momentum space makes sense, 
it does not in configuration space where wave functions imply an integral over
$p$  of momentum wave functions together with spherical Bessel functions of
different $l$. A possible comparison could involve the integral of the different
momentum wave functions with the same spherical Bessel function $j_1(pr)$ for
the lowest pseudospin doublet (or $j_2(pr)$ for the second one).
We defined these quantities as: 
\begin{equation}
f(r)= \sqrt{\frac{2}{\pi}}\; r \int dp\; f(p)\; j_1(pr) \;({\rm or}\; j_2(pr))\,,
\label{eq:fnr}
\end{equation}
which, using Eq. (\ref{eq:normp}), are found to verify the normalization 
condition:
\begin{equation}
\int dr\; f^2(r) =1\,.
\label{eq:normr}
\end{equation}
Interestingly, the quantities $f(r)$ together with the associated angular
momentum structure corresponding to the states  $p_{3/2},\; p_{1/2}$  
(or $d_{5/2},\;d_{3/2}$), show some similarity with the small components 
in the Dirac mean-field phenomenology. It is reminded that the examination 
of the equation allowing one to determine these components provided some 
justification for the existence of pseudospin doublets \cite{Gino97,Gino98}.
Taking into account Eq. (\ref{eq:relp}), the functions $f(r)$
are expected to verify the following relation in the pseudospin-symmetry limit:
\begin{eqnarray}
f_{1d_{3/2}}(r)=f_{1s_{1/2}}(r)=f_{1p_{1/2,3/2}}(r)\,,
\nonumber \\
f_{1f_{5/2}}(r)=f_{1p_{3/2}}(r)=f_{1d_{3/2,5/2}}(r)\,.
\label{eq:relr}
\end{eqnarray}
Results showing the $f(r)$ functions are presented in Figs.
\ref{psr1} and \ref{psr2}, in full correspondence with those shown in
Figs. \ref{psp1} and  \ref{psp2} for momentum space. 
Qualitatively, the different curves evidence a similar pattern. 
There is not much sensitivity to the interaction model 
and the slight sensitivity to the pseudospin doublet can be
mainly ascribed to the corresponding binding, 
the results for the less bound system tending to extend to larger radial
distances. 

In each panel, curves corresponding to the pseudospin assignment made here have
the same sign within the largest range of their contribution to the
normalization given by Eq. (\ref{eq:normr}). They evidence some spreading 
but fall slightly apart on one side and on the other side of the expected result 
in the pseudospin-symmetry limit. The curves corresponding to the 
standard pseudospin assignment evidence a striking difference. One of them
changes sign within the range of its main contribution to the normalization.
Thus, these results show that, like in momentum space, if one considers 
only the similarity of the $f(r)$ functions of the PS partners,
it is more natural 
to consider the states $1d_{3/2},\; 1s_{1/2}$  (or $1f_{5/2},\;1p_{3/2}$),  
as a pseudospin doublet rather than the states $1d_{3/2},\; 2s_{1/2}$  
(or $1f_{5/2},\;2p_{3/2}$). The argument will get further strength 
if we notice that the above spreading represents for a large part a difference
in the binding energy or in the momentum distribution of the wave functions. 
It is found that Eq. (\ref{eq:relr}) is much better fulfilled than 
what Figs. \ref{psr1} and \ref{psr2} suggest if we require that 
the average square momentum be the same, as expected in the
pseudospin-symmetry limit. This can be checked by using scaled wave functions.
\section{Relation with results  from a relativistic approach}
Since the PS symmetry has been considered in the last decade as a symmetry with
deep roots in the relativistic theory of atomic nuclei \cite{Gino97}, 
arises the question of whether the above results can be cast into 
this approach, which can be described by the following equations:
\begin{eqnarray}
&&\vec{\sigma} \!\cdot\! \vec{p}\;\;\psi_{<}(\vec{r}) -(E-M-V(r))\;\psi_{>}(\vec{r}) =0,
\nonumber \\
&&\vec{\sigma} \!\cdot\! \vec{p}\;\;\psi_{>}(\vec{r})- (E+M-\Delta(r))\;\psi_{<}(\vec{r}) =0.
\label{eq:dirac}
\end{eqnarray}
where $\psi_{>}(\vec{r}) $ and $\psi_{<}(\vec{r})$ represent, respectively, 
the usual large and small component of the Dirac spinor. 
Various models are considered in the following.

\subsection{Model inspired from the non-relativistic one}
\label{ssec:rel1}
Generally, the potentials $V(r)$ and $\Delta(r)$ are taken 
as the sum and the difference of the scalar and vector potentials, 
$V(r)=V_v(r)+V_s(r)$ and $\Delta(r)=V_v(r)-V_s(r)$, but , formally, 
nothing prevents one to take $V(r)$  as given by Eq. (\ref{eq:vnr}) 
and $\Delta(r)$ as a constant. This model has not much to do with the usual
relativistic mean-field phenomenology but it can be easily studied. 
Actually, it does not add relevant information to the non-relativistic model 
studied in the previous section. 
Very similar results are obtained for the spectrum as well as
for the wave function as far as the large component is concerned.

\subsection{Model inspired from the Chen {\it et al.} one}
A more interesting comparison involves the model considered 
by Chen {\it et al.} \cite{Chen02,Chen03}. Apart from the energy scale, 
the pseudospin-symmetry spectrum shown 
in Fig. \ref{fig_ho} evidences a striking similarity with theirs. 
There could be also some
similarity with results obtained by Lisboa {\it et al.} \cite{Lis04}, 
which are essentially the same as those obtained by  Chen {\it et al.}, 
but the authors modified the assignment of the quantum number $n$ 
to make it identical to the number of nodes of the radial wave functions. 
The authors compare the results in the cases of the two symmetry limits, 
arguing  about missing and intruder states. 
The question therefore arises whether there is a continuity between the two
limits, as observed in the non-relativistic case discussed in previous
sections. Such a study could also help in clarifying the assignment 
of the $n$ quantum number.

In order to study the relationship between the two symmetry cases, 
one has to determine the two potentials,  $V(r)$ and $\Delta(r)$, 
that enter 
the following standard equations for the large and small components, 
$G(r)$ and  $F(r)$:
\begin{eqnarray}
&&\hspace*{-1.3cm}\Bigg[\!-\frac{d^2}{dr^2}\!+\!\frac{l(l\!+\!1)}{r^2}
-(E\!+\!M\!-\!\Delta(r)) \, (E\!-\!M\!-\!V(r))\Bigg]\,G(r) -
\frac{\frac{d\Delta(r)}{dr} (\frac{d}{dr}\!+\!\frac{\kappa}{r},\,G(r)) }{  
E\!+\!M\!-\!\Delta(r)} =0\, , 
\label{eq:GG} \\
&&\hspace*{-1.3cm}\Bigg[\!-\frac{d^2}{dr^2}\!+\!\frac{\tilde{l} (\tilde{l}\!+\!1)}{r^2}
-(E\!+\!M\!-\!\Delta(r)) \, (E\!-\!M\!-\!V(r))\Bigg]\,F(r) -
\frac{\frac{dV(r)}{dr} (\frac{d}{dr}\!-\!\frac{\kappa}{r},\,F(r)) }{ 
E\!-\!M\!-\!V(r)}=0\, , 
\label{eq:FF}
\end{eqnarray}
where the notation such as $(\frac{d}{dr}\!+\!\frac{\kappa}{r},\,G(r))$ 
stands for $(G'(r)\!+\!\frac{\kappa}{r}\,G(r))$. In the cases of spin- and pseudospin-symmetry limits, 
the potentials $V(r)$ and $\Delta(r)$  should be respectively given by:
\begin{eqnarray}
&V(r)=V_0(r)=\frac{1}{2}\omega^2M r^2, &\hspace*{1.6cm}\Delta(r)=0\,,
\nonumber \\
&V(r)=0, &\hspace*{1cm}\Delta(r)=\Delta_0(r)=\frac{1}{2}\omega^2M r^2\,.
\end{eqnarray}

A first guess to determine the transition potential would consist 
in multiplying $V_0(r)$ and $\Delta_0(r)$ by the factors $1-x$ and $x$, 
and allow $x$ to vary between 0 and 1. It can be however shown that solutions 
of Eqs. (\ref{eq:GG}, \ref{eq:FF}) have then an oscillatory behavior at large $r$, 
which complicates their determination and could obscure conclusions. 
As we are interested in finding at least one way to ensure a continuous transition
between the two symmetry limits, we proceed differently. 
Instead of fixing {\it a priori} the transition potential, we determine this
potential from minimal consistency conditions. To illustrate the method 
we adopted, we consider here with some detail the case  $n=1$ 
and $l=\tilde{l}+1$.

First, as  the above mentioned oscillatory character of solutions is due to
the appearance of an attractive $r^4$ term in the product 
$(M\!+\!E\!-\!\Delta(r)) \, (M\!-\!E\!+\!V(r))$, we skip this term so that the
remaining part behaves as an ordinary harmonic-oscillator potential.
This product thus read: 
\begin{eqnarray}
(E\!+\!M\!-\!\Delta(r)) (E\!-\!M\!-\!V(r))=E^2\!-\!M^2-\frac{M\omega^2r^2}{2}
\Big((1\!-\!x)(E\!+\!M) +x (E\!-\!M)   \Big)\,.
\label{eq:relVD}
\end{eqnarray}
It allows one to determine $V(r)$ once $\Delta(r)$ has been determined 
from other sources. Second, the examination  of the last term in 
Eq. (\ref{eq:GG})  in the pseudospin-symmetry limit ($V(r)=0$) 
shows that the singular behavior due to its denominator is exactly 
cancelled by the action on  $G(r)$ of the operator, 
$\frac{d}{dr}\!+\!\frac{\kappa}{r}$, at the numerator. The equation 
can then be solved analytically, allowing one to recover the solution 
which is known in any case. It is not difficult to see that the procedure 
can be extended to any $x$ in the simplest case of orbits with $n=1$ 
and $l=\tilde{l}+1$ 
(states $1p_{1/2},\;1d_{3/2},\;1f_{5/2},\;1g_{7/2}, \cdots $).
The solution assumes the form:
\begin{eqnarray}
G(r) \propto r^{(l+1)} \,{\rm exp}(-\alpha^2\,r^2/2)\, ,
\end{eqnarray}
while the last term in Eq. (\ref{eq:GG}) can be written as:
\begin{eqnarray}
-\frac{\frac{d\Delta(r)}{dr} (\frac{d}{dr}\!+\!\frac{\kappa}{r},\,G(r))}{E\!+\!M\!-\!\Delta(r)}
=-\frac{\frac{d\Delta(r)}{dr}\;(2l\!+\!1\!-\!\alpha^2\,r^2)}{(E\!+\!M\!-\!\Delta(r))\;r}
 \,G(r)=-x\, \frac{(2l\!+\!1) \,\omega^2M}{E\!+\!M} \, G(r)\,.
\label{eq:compac}
\end{eqnarray}
By inserting the above expression in Eq. (\ref{eq:GG}), 
one gets the following relations to be fulfilled:
\begin{eqnarray}
&&\hspace*{-0.5cm}\alpha^2-\omega\,
\sqrt{\frac{M }{2} \Big((1\!-\!x)(E\!+\!M) +x (E\!-\!M)   \Big)}=0\,, 
\label{eq:alpha}
\\
&&\hspace*{-0.5cm}E^2\!-\!M^2 - \omega \,(2l\!+\! 3)\;
\sqrt{\frac{M }{2} \Big((1\!-\!x)(E\!+\!M) +x (E\!-\!M)   \Big)} 
+x \frac{(2l\!+\!1)\,\omega^2M}{E+M}=0\, .
\label{eq:compag}
\end{eqnarray}
The first of them is obtained from cancelling the coefficient 
of the highest $r$-power term in the equation determining $G(r)$ 
(term proportional to $r^{l+3}$). The second one allows one to get the
energy and is obtained from cancelling the next term 
in the same equation (term proportional to $r^{l+1}$). 
It can be checked that the states 
$1p_{1/2},\;1d_{3/2},\;1f_{5/2},\;1g_{7/2}, \cdots $ in the spin-symmetry 
limit are in a one-to-one correspondence with the same states in the 
pseudospin-symmetry limit. Assigning to these states the same $n$ 
quantum number as for the spin-symmetry case, it is found that this number
coincides at $x=1$ with the one assigned by Chen {\it et al.}, in this case 
from algebraic considerations \cite{Chen02}.

For completeness, we give the potentials $\Delta(r)$ 
and $V(r)$. The potential $\Delta(r)$ is obtained by solving the equation 
that results from the last equality in Eq. (\ref{eq:compac}):
\begin{eqnarray}
-\frac{\frac{d\Delta(r)}{dr}}{E\!+\!M\!-\!\Delta(r)}
=-x\,\;\frac{\omega^2M}{E\!+\!M}\;
\frac{(2l\!+\!1)\;r}{(2l\!+\!1\!-\!\alpha^2\,r^2)} \, .
\label{eq:compad}
\end{eqnarray}
The  solution of this equation is given by: 
\begin{eqnarray}
E\!+\!M\!-\!\Delta(r)= \;(E\!+\!M)\;
\Big(1\!-\!\frac{\alpha^2\,r^2}{2l\!+\!1}\Big)^{ 
\frac{x (2l\!+\!1)\omega^2 M }{2\alpha^2 (E\!+\!M)}}\,.
\label{eq:compae}
\end{eqnarray}
In integrating Eq. (\ref{eq:compad}), we fixed the integration constant by
requiring that  $\Delta(r)=0$, as expected in the  spin-symmetry limit
corresponding to $x=0$, hence  the overall factor, $E\!+\!M$, at the r.h.s. 
of  Eq. (\ref{eq:compae}).  The other factor is well defined 
when the quantity, $2l\!+\!1\!-\!\alpha^2\,r^2$,  is positive. 
For negative values, it is appropriate to replace this factor by 
$e^{\pm i \pi}(\alpha^2\,r^2\!-\!2l\!-\!1)$. The imaginary phase 
that could then occur, 
exp$(\pm i\pi\frac{x (2l\!+\!1)\omega^2 M }{2\alpha^2 (E\!+\!M)})$,  
can be removed but should be accounted for in
determining the small component. 
It can be checked that the exponent is equal to 1 at $x=1$, 
so that the factor at the denominator in Eq. (\ref{eq:compac}) 
is exactly cancelled by the factor $(2l\!+\!1\!-\!\alpha^2\,r^2)$ 
at the numerator (up to a constant factor). The expression of the small
component is easily obtained and reads:
\begin{eqnarray}
F(r)=\frac{ (\frac{d}{dr}\!+\!\frac{\kappa}{r},\,G(r))}{E\!+\!M\!-\!\Delta(r)}
= \frac{2l\!+\!1}{E\!+\!M}\;
\Big(1\!-\!\frac{\alpha^2\,r^2}{2l\!+\!1}\Big)^{(1- 
\frac{x (2l\!+\!1)\omega^2 M }{2\alpha^2 (E\!+\!M)})}\;\;\frac{G(r)}{r}\, .
\label{eq:spect4f}
\end{eqnarray}
The potential $V(r)$ can be obtained from the expression of $\Delta(r)$, given
by Eq. (\ref{eq:compae}), and from Eq. (\ref{eq:relVD}),  
which can be cast into the form:
\begin{eqnarray}
(M\!+\!E\!-\!\Delta(r))\; (M\!-\!E\!+\!V(r))=M^2\!-\!E^2 +\alpha^4 \, r^2\,.
\label{eq:compah}
\end{eqnarray}
The above approach can be extended to other orbits, either with non-zero $n$ 
or with $\tilde{l}=l\!+\!1$.  We give below equations allowing one 
to determine $E$ as a function of $x$ and make some comments 
about obtaining these results. We refer to the appendix \ref{app:A} 
for further details.

In all cases, we assume that the wave function 
writes as the product of an exponential factor, ${\rm exp}(-\alpha^2r^2/2)$, 
and a polynomial term, similarly to the harmonic oscillator wave functions 
that are recovered at $x=0$ for the large component, $G(r)$, 
or at $x=1$ for the small one,  $F(r)$. The consideration 
of the highest r-power term in Eq. (\ref{eq:GG})  ensures that the expression 
of $\alpha^2$  has always the form given by Eq. (\ref{eq:alpha}), while  
the consideration of the lowest r-power term in the equation requires 
that the wave function contains the minimal well known factor $r^{l\!+\!1}$. 

For $\tilde{l}=l\!-\!1$ ($\kappa=l$) and any $n$ 
($np_{1/2},\;nd_{3/2},\;nf_{5/2},\;ng_{7/2}, \cdots $), we assume 
that the polynomial part of the wave function has a structure similar 
to the spin-symmetry case and, 
in particular, has the same number of nodes. From considering the next 
to the highest r-power term ($r^{2n\!+\!l\!-\!1}$) in Eq. (\ref{eq:GG}), one gets 
the equation giving $E$  as a function of $x$:
\begin{eqnarray}
E^2\!-\!M^2 - (4n\!+\!2l\!-\! 1)\;\alpha^2
+x \,\frac{(4n\!+\!2l\!-\!3)\,\omega^2M}{E+M}=0 \, .
\label{eq:specmn2}
\end{eqnarray}
The lowest r-power terms are useful in determining the polynomial part 
and, from it, the expression of the potential $\Delta(r)$. 
This is tractable for the case $n=1$ considered above (first-degree equation) 
and for the case $n=2$ detailed in the appendix \ref{app:A} (second-degree equation) 
but the difficulty of the task quickly increases with $n$ ($n$-degree equation).
\begin{figure}[htb]
\mbox{ \epsfig{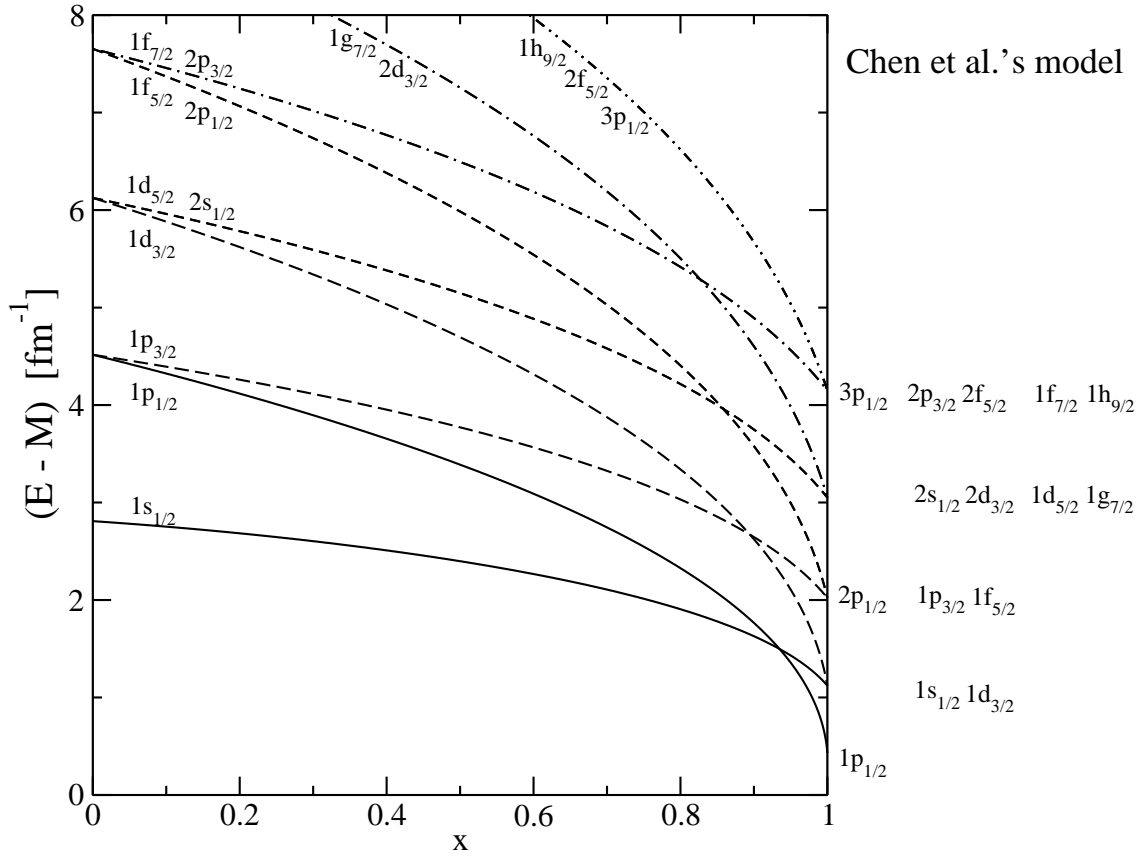}}
\caption{Single-particle spectrum in a relativistic approach 
with a harmonic-oscillator inspired interaction: transition from the spin-symmetry limit ($x=0$)
to the pseudospin-symmetry limit ($x=1$). We limited the representation of
states to those corresponding to the complete multiplets shown on the r.h.s. 
of the figure. Some higher-energy states are therefore not shown. 
\label{fig_horel}}
\end{figure}  

For finding solutions for the case with $\tilde{l}=l\!+\!1$ 
($\kappa=-(l\!+\!1)$) 
and any $n$ ($ns_{1/2},\;np_{3/2},\;nd_{5/2},\;nf_{7/2}, \cdots $), 
we notice that there is some symmetry between Eq. (\ref{eq:GG}) 
and Eq. (\ref{eq:FF}). The orbital angular momentum $l$  is replaced by
$\tilde{l}$ and the solution for the small component has the same number of
nodes at $x=0$ and $x=1$. What we did previously for the large component can
therefore be repeated here for the small one with the appropriate replacements 
($\Delta(r) \rightarrow V(r)$, $M  \rightarrow -M $,  $\kappa  \rightarrow
-\kappa $, $x \rightarrow 1-x$).
Thus Eqs. (\ref{eq:compad}) and (\ref{eq:compae}) for $n=1$ become:
\begin{eqnarray}
-\frac{\frac{dV(r)}{dr}}{E\!-\!M\!-\!V(r)}
=-(1-x)\,\;\frac{\omega^2M}{E\!-\!M}\;
\frac{(2\tilde{l}\!+\!1)\;r}{(2\tilde{l}\!+\!1\!-\!\alpha^2\,r^2)} \, ,
\label{eq:compadd}
\end{eqnarray}
and: 
\begin{eqnarray}
E\!-\!M\!-\!V(r)= \;(E\!-\!M)\;
\Big(1\!-\!\frac{\alpha^2\,r^2}{2\tilde{l}\!+\!1}\Big)^{ 
\frac{(1\!-\!x) (2\tilde{l}\!+\!1)\omega^2 M }{2\alpha^2 (E\!-\!M)}}\,.
\label{eq:compaed}
\end{eqnarray}
The general equation allowing one to determine the spectrum for any $n$ 
now reads:
\begin{eqnarray}
E^2\!-\!M^2 - (4n\!+\!2\tilde{l}\!-\! 1)\;\alpha^2
+(1\!-\!x) \, \frac{(4n\!+\!2\tilde{l}\!-\!3)\,\omega^2M}{E-M}=0 \, .
\label{eq:specmn4}
\end{eqnarray}

The results for the spectrum are shown in Fig. \ref{fig_horel} for $x$ varying 
from 0 to 1. They have been obtained using parameters identical to
those used in Refs. (\cite{Chen03}, \cite{Lis04}) ($M=10$ fm$^{-1}$ 
and $\omega=2$ fm$^{-1}$),
so that to facilitate some comparison with their results. This choice 
also offers the advantage that the spectrum at $x=1$ is less compressed as for
the choice  $\omega=1$ fm$^{-1}$ used in an earlier work \cite{Chen02}.
Examination of the figure shows that there is an one-to-one correspondence
between states in the spin- and the pseudospin-symmetry limits, indicating  
that there is no missing states as one goes from some limit to the other one.
\begin{figure}[htb]
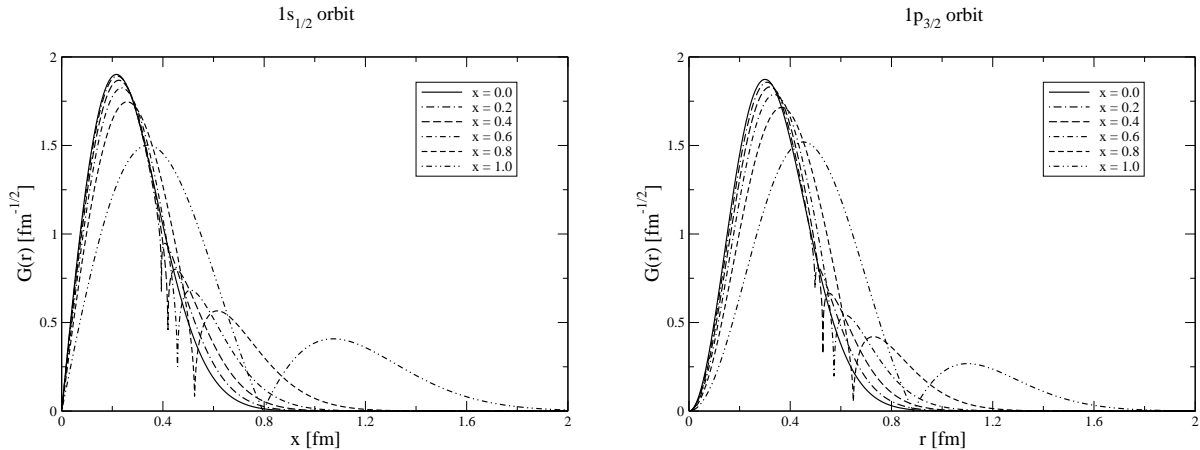

\mbox{\epsfig{ file=fns0rb.eps, width=7.5cm}\hspace*{0.7cm}
\epsfig{ file=fns1rb.eps, width=7.5cm}}
\caption{Appearance of an extra node in the component $G(r)$ when going 
from the spin- to the pseudospin-symmetry limit for orbits $1s_{1/2}$ 
and $1p_{3/2}$ (left and right panels respectively) 
with the Chen {\it et al.} model. The normalization assumes the relation 
$\int dr \, (G^2(r)+F^2(r))=1$. \vspace*{1.0cm}
\label{nodech}}
\end{figure}  

In relation with the wave functions, we recover their expressions expected 
in the two symmetry limits. We find, in particular, that the large component 
of the states with $\tilde{l}=l\!+\!1$, corresponding to a given $n$ 
(as defined here by continuity with the one for the spin-symmetry case), 
acquires an extra node while one goes from the spin- 
to the pseudospin-symmetry limit.
It shows that establishing a correspondence of the states with the same number 
of nodes in both limits, as done in Ref.  \cite{Lis04}, 
could lead to misleading conclusions. 
Actually, examination of the wave functions shows that the extra node appears 
at a distance where the contribution to the norm is almost saturated. 
This result is similar to the one obtained in the previous section 
for a non-relativistic model but the rate of saturation is smaller 
in the present case. Contributions to the norm beyond the last node 
amount to  around 10\% for the $1s_{1/2}$ state, 
7\% for the  $1p_{3/2}$ state and less for higher angular momenta. 
Some results are shown in Fig.
\ref{nodech}  for the $1s_{1/2}$ and  $1p_{3/2}$ states while expressions for
the wave functions can be found in the appendix \ref{app:A}. 
Contrary to the nonrelativistic case, the appearance of an extra node is not
progressive. Virtually, the node is present at $x=0$ but it does not show up
because the factor responsible for it, $1-\frac{\alpha^2\,r^2}{2l+1}$, 
is taken with the power $0$, making this factor equal to 1 for any $r$ value. 
This is a particular feature of the transition model we used.
The appearance of an extra node also holds for the small component, 
but only for the states with $\tilde{l}=l\!-\!1$ and when making 
the inverse path.

We assumed in the above study of the continuity between the spin- 
and pseudospin symmetry limits that the last term in  Eq. (\ref{eq:GG}) 
was linear  in $x$, partly because it was the simplest possible assumption 
to start with. Many other assumptions could be made. A particularly attractive
one for the case $\tilde{l}=l\!-\!1$ supposes to replace the factor  
$-x\, \frac{(2l\!+\!1) \,\omega^2M}{E\!+\!M} $ at the r.h.s. of 
Eq. (\ref{eq:compac}) by  $-2x\frac{(2l\!+\!3\!-\!2x) \,\alpha^4}{E^2\!-\!M^2}$.
The dependence on $x$ of this expression is more complicated than
for the choice retained here ($\alpha$ also depends on $x$), but the result for
the relation between $E$ and $x$ turns out to be formally simpler. 
It is given by:
\begin{eqnarray}
E^2\!-\!M^2= (4n\!+\!2l\!-\!1\!-\!2x)\alpha^2\, .
\end{eqnarray}
In writing the above result, we anticipated its generalization to any $n$.
A similar result could be obtained for the case $\tilde{l}=l\!+\!1$ with the
replacements of $l$ by $\tilde{l}$ and $x$ by $1-x$. It reads: 
\begin{eqnarray}
E^2\!-\!M^2= (4n\!+\!2l\!-\!1\!+\!2x)\alpha^2\, .
\end{eqnarray}
 The continuous
transition between the spin-and pseudospin-symmetry limits is somewhat
straightforward from the above expressions. In both cases, 
the result should be completed by the derivation of the potentials 
$V(r)$ and $\Delta(r)$, which is not necessary simpler than before.

\subsection{Model with $V(r)$ and  $\Delta(r)$ inspired from Woods-Saxon ones}
In previous sections, we examine models  that were providing  some continuity
between the spin- and pseudospin-symmetry limits. There is no guarantee that
such a continuity exists in all cases however. On the one hand, some states may
disappear and reappear (or appear and disappear) in going from one limit to the
other but in cases considered up to now we could find a model avoiding this
drawback. On the other hand, some states may disappear (or appear), 
what cannot be excluded in a relativistic model where the spectrum 
is not bounded from below. A reason to be suspicious comes from the fact 
that realizing the pseudospin symmetry with a Woods-Saxon type potential 
supposes that its depth is larger than twice the nucleon mass, which is the
obvious sign of a relativistic regime.
\begin{figure}[htb]
\rotatebox{ -90}{\epsfig{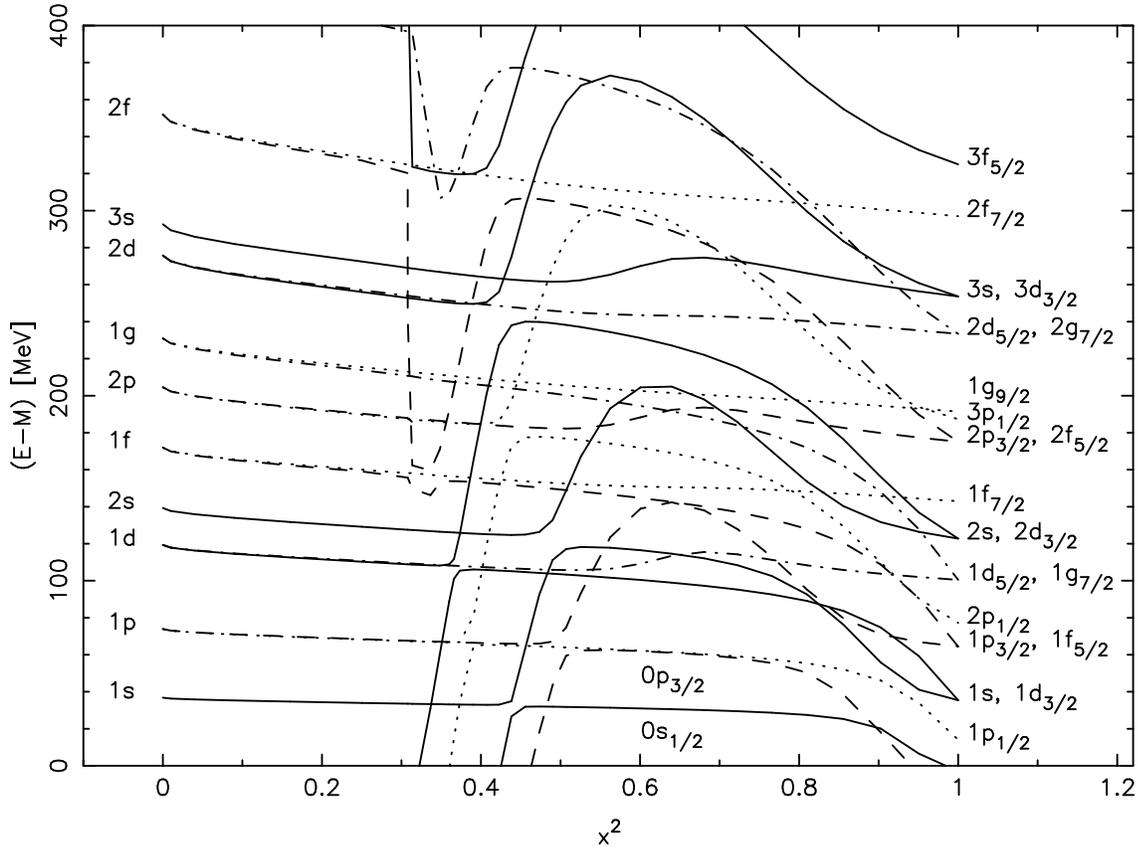}}
\caption{Single-particle spectrum in a relativistic approach with a Woods-Saxon
inspired interaction: transition from the spin-symmetry limit ($x=0$)
to the pseudospin-symmetry limit ($x=1$). Notice that our conventions for the
potentials at  $x=0$ and $x=1$ imply positive values for the quantity, $E-M$,
represented on the vertical axis (see appendix \ref{app:B}).
\label{fig_hwsrel}}
\end{figure}  

We examined various transition models from the spin-to the pseudospin-symmetry
limits but, due to various problems (numerics, interpretation), we could not get
clear conclusions in most cases. We present here results for one case where some
interpretation is possible. Moreover, they provide interesting clues as for what
could occur between the two symmetry limits. The model is described in the
appendix while the spectrum is shown in Fig. \ref{fig_hwsrel}. 
An essential difference with  Chen {\it et al.}'s model is that the
potentials at  $x=0$ and $x=1$, instead of going to $\infty$ with $r$, 
are now limited by an upper finite value.

We first notice that the spectrum in the spin-symmetry limit looks very much
like the usual one. As far as the order of the states is concerned, 
the spectrum in the pseudospin-symmetry limit is similar to the one 
for the WS model shown in Fig. \ref{fig_ws}. 
It would also be similar to the one obtained by Chen {\it et al.} \cite{Chen03}, 
had the authors chosen to raise the degeneracy of their HO interaction model 
with a Woods-Saxon type correction having the opposite sign. 
Considering the large component of the wave functions, it is found that
all states with $j=l-1/2$ and same  $n$ quantum number (defined here 
by the order of appearance in the spectrum) have the same number of nodes 
in the pseudospin-symmetry limit as in the spin-symmetry limit. Instead, 
states with  $j=l+1/2$ get an extra node, located however in a domain
where the contribution to the norm beyond this node is, generally, small. 
This observation is similar to the one we could make for the non-relativistic 
case or for  Chen {\it et al.}'s model \cite{Chen02}.  

At first sight, from the qualitative agreement of the spectra in the SS and PSS
limits with those obtained previously in this work, we could expect similar
conclusions for the spectrum between these two limits. 
Looking at this one however reveals important differences. 
It is found that there is continuity between $x=0$ and $x=1$ for all states 
with $j=l+1/2$, without change for the $n$ quantum number 
(always defined here by the order of appearance in the spectrum).
This is also satisfied for states with  $j=l-1/2$ and $l \geq 3$ 
but the picture differs for states $np_{1/2}$ and $nd_{3/2}$. 
As these states evolve from the spin- to the pseudospin-symmetry limit, 
their  $n$ quantum number increases by one unit. 
This change is related to the appearance 
of an extra node in the core of the large component of the wave function. 
As a counterpart of the change in $n$, the states  $1p_{1/2}$ and $1d_{3/2}$ 
in the pseudospin-symmetry limit cannot be related to any state 
in the spin-symmetry limit. As seen in the figure, they disappear from the
spectrum around $x=0.5-0.6$, where the interaction is strongly modified. 
This fact explains the disappearance of the PS partner 
of the intruder $1s_{1/2}$ state. It is also seen on the figure 
that a new kind of states
with $j=l+1/2$ (to whom we assign, arbitrarily, $n=0$) appears 
and disappears between $x=0$ and $x=1$, without spoiling
the symmetry properties of the spectra obtained in these limits.

One can think, perhaps, that the extra states that appear with $E > 0 $ 
for values of $x > 0$ are simple artefacts of the interaction model. 
However, among them, those states with $j=l-1/2$ are necessary in order 
the PS symmetry be satisfied for  $x =1$. The other states with $j=l+1/2$ 
are not required by this symmetry but 
their appearance strongly influences the spectra of the states 
that are above, pushing them upwards and evidencing level-repulsion properties 
around $x=0.5- 0.7$. Level-repulsion properties are also observed for the levels
with $j=l-1/2$ in the same range. In these cases, this is the range where the
wave function can be strongly changed and acquires an extra node. In some of
these cases, the small and large components become comparable, suggesting the
role of relativity in this change. It is noticed that, if we ignore 
the small region where the level-repulsion effects for these states appear, 
the spectra look continuous from
the spin- to the pseudospin-symmetry limits. Properties of the large component
of the wave function before and after the pseudo-crossing point with a similar
energy are not much affected. 
These features are  similar to the ones for the dependence of
single-particle spectra as a function of the deformation of the nucleus. 
   
The detail of the above results depends on the transition model but part of it, 
like the appearance of an extra node as $x$ increases as well as the appearance 
(or disappearance) of some states from the positive part of the SP  spectrum, 
may point to relevant features of the models used to
describe the interaction in the spin- or pseudospin-symmetry limit. 
Further studies could be useful to determine, in particular, 
whether the disappearance of states could be completely removed or, 
on the contrary, be extended to all states with $j=l-1/2$ and $n=1$, 
then providing support for the missing partners of intruder states. 
Whatever the picture,  the results point to an important 
breakdown of the pseudospin symmetry. 
States disappear from the spectrum in one case or, in the other case,  
the assignment of states to pseudospin doublets differs from the one usually
assumed, while agreeing with results from other parts of this work.

\section{Conclusions}
In this paper, we examined whether there could be some continuous transition
between the spin-symmetry and the pseudospin-symmetry limit description 
of atomic nuclei and looked at what would be the implications of such a result.
This was done  by introducing an interaction that is a continuously varying 
superposition  of potentials satisfying these symmetry properties.

Within a non-relativistic approach, we found that there could be a one-to-one
correspondence between the spectra calculated in the SS and PSS limits. 
This is obtained with an interaction that is non-local in the PSS limit,
resulting in a state ordering of the spectrum different from the one 
usually expected (the ground state is not necessarily a $s_{1/2}$ one).
For the transition model we considered, energy shifts are larger than 
for the spin-orbit splitting but they do not generally exceed the splitting
corresponding to two major shells. The results thus suggest to associate 
in the same PS doublet states $1s_{1/2}$ and  $1d_{3/2}$,  
$1p_{3/2}$ and  $1f_{5/2}$, $\cdots$, rather than states $2s_{1/2}$ 
and  $1d_{3/2}$,   $2p_{3/2}$ and  $1f_{5/2}$, $\cdots$. Our suggestion is
further supported by the analysis of wave functions which leads to results 
closer to the PS-limit expectations with our assignment than with the usual
one.  With this respect, we notice that wave functions for the states $1s_{1/2}$, 
$1p_{3/2}$,  $\cdots$, get an extra node in the PSS limit but this node is
located in the tail of the wave function rather than in its bulk as for states 
$2s_{1/2}$, $2p_{3/2}$,  $\cdots$. This prevents one to identify these states
together, what is further supported by our study looking at a continuous
transition from the SS to the PSS limit. 

We also looked at a possible continuous transition  between the spectra 
calculated in the SS and PSS limits within a relativistic approach 
as this one was thought to provide some support for PSS doublets. 
Without regard to the origin of the interaction, we first notice 
that previous results can be cast into the relativistic approach 
by identifying the potential $V(r)$ to the one  used in the non-relativistic 
case and taking $\Delta(r)$ as a constant. 
Another relativistic model is the
one considered by Chen {\it et al.} \cite{Chen02}, based on oscillator
type potentials. In this case, we were able to find a potential allowing one to
get a continuous transition between the SS and PSS spectra, which turn out to be
very similar to those we get in the non-relativistic case. Contrary to
statements made by the authors, there is no need to include intruder or missing
states to make the spectra consistent with each other. We simply have to accept
that the order of the states in the PSS limit differs from the usual one. 
With the intent to study more realistic models, 
we finally consider the case of finite-size potentials. Apart from the energy
scale, the spectra in the SS and PSS limits were found to be similar 
to those obtained in the previous cases, suggesting that one could establish some
continuity between them as done previously. 
However, while all states in the SS limit could be
continuously related to states in the PSS limit, the reciprocal was not
true. Some states in the PSS limit
could not be related to states in the SS limit and, actually,
were found to disappear from the positive-energy spectrum.
Interestingly, the spectrum so obtained is found to be similar for a part 
to the observed one, including the assignment for PSS doublets. 
Thus, these results could throw light, in particular, about the disappearance 
of the PS partners of the intruder states when one goes from the
PSS to the SS limit. 
The results nevertheless call for two important remarks. 
First, as the disappearance implies a limited number of states 
with $n=1$ and $j=l-1/2$, there is likely a dependence of the results 
on the interaction models we used in the SS and PSS limits. 
Second, the disappearance of states from the positive-energy spectrum 
points to a very strong violation of the PS symmetry in this case, 
more important 
than the one found in the other approaches where only a re-ordering 
of the levels was observed. Moreover, the assignment for PS doublets 
then implies that the large component of the wave function of one 
of the partners gets an extra node in its bulk when going 
from the SS to the PSS limit. This feature represents another violation 
of the PS symmetry.

From examining wave functions, it was shown in the literature that the
pseudospin symmetry could not be as good as the spin one 
\cite{Mar04,Mar05b,Mar08}, 
pointing to the accidental character of the validity of this symmetry to
describe the SP-level spectrum.  
Starting from a different viewpoint, the present results largely support this
observation. If some doubt is allowed, it is due to one of the studies 
we performed within a relativistic framework. In this case, however, further studies 
could be useful to determine in what extent the results depend 
on specific features of the interaction we used. Either these results are
confirmed and they imply, intrinsically, a very strong violation 
of the pseudospin symmetry, or, they confirm results of the other approaches, 
with a  violation of this symmetry that is smaller but nevertheless larger 
than the spin-symmetry one.

\noindent
{\bf Acknowledgments}\\
This work has been supported by the MEC grant FIS2005-04033.

\appendix
\section{Transition from the SS to PSS limit in  the Chen {\it et al.} model}
\label{app:A}
We give in this appendix some further details relative to the calculation 
of the spectrum between the spin- and pseudospin-symmetry limits 
in the  Chen {\it et al.} model \cite{Chen02}. 
They successively concern the equations relating $E$ to $x$, and the parameters 
 $\omega$ and $M$,
the determination of the transition potential for the case  $n=2$ 
and $\tilde{l}=l\!-\!1$, and, somewhat sketched, for the case $n=1$ 
and $\tilde{l}=l\!+\!1$.

\noindent
$\bullet$ {\it Remark about the relation of $E$ to $\omega$}\\ 
It is first noticed that the equations allowing one to determine the relation 
of $E$ to $x$, Eqs. (\ref{eq:specmn2}) and (\ref{eq:specmn4}), 
are quadratic in $\omega$ and, therefore, apparently  differ from those 
given in Ref. \cite{Chen02} in the case $x=1$ and $x=0$ respectively, 
which are linear in $\omega$. 
They can nevertheless be rearranged so that to express $\omega$ in terms 
of  $E$ and $x$. When this is done, one recovers the expressions given 
in the above work:
\begin{eqnarray}
&&\omega\,\Big(2n\!+\!l\!-\!\frac{3}{2}\Big)=\omega\,\Big(2n\!+\!\tilde{l}\!-\!\frac{1}{2}\Big) 
=(E\!+\!M)\,\sqrt{\frac{E\!-\!M}{2M}}\hspace*{1cm} ( \tilde{l}=l\!-\!1,\; x=1)\,,
\\
&&\omega\,\Big(2n\!+\!\tilde{l}\!-\!\frac{3}{2}\Big)=\omega\,\Big(2n\!+\!l\!-\!\frac{1}{2}\Big) 
=(E\!-\!M)\,\sqrt{\frac{E\!+\!M}{2M}}\hspace*{1cm} ( \tilde{l}=l\!+\!1,\;x=0)\,. 
\end{eqnarray}
%


\noindent
$\bullet$ {\it Case of states with $n=2$ 
and $j=l-1/2$: $2p_{1/2},\;2d_{3/2},\;2f_{5/2},\cdots$, ($\tilde{l}=l\!-\!1$)}\\
We assumed 
that the wave function keeps the same form as for $x=0$ or $x=1$, 
$G(r) \propto r^{(l+1)}(\alpha^2\,r^2 -y) \,{\rm exp}(-\alpha^2\,r^2/2)$, 
which implies that the last term in  Eq. (\ref{eq:GG}) can be written as:
\begin{eqnarray}
-\frac{\frac{d\Delta(r)}{dr} (\frac{d}{dr}\!+\!\frac{\kappa}{r},\,G(r))}{E\!+\!M\!-\!\Delta(r)}
&=&-\frac{\frac{d\Delta(r)}{dr}\;
\Big(\alpha^2\,r^2(2l\!+\!3\!+\!y)-(2l\!+\!1)\,y -\alpha^4\,r^4\Big) }{ 
(E\!+\!M\!-\!\Delta(r))\,(\alpha^2\,r^2 -y)\;r} \,G(r)
 \nonumber\\
&=&-x\, \frac{(2l\!+\!5) \,\omega^2M}{E\!+\!M} 
\frac{(\alpha^2\,r^2 -z)}{(\alpha^2\,r^2 -y)}\, G(r)\,.
\label{eq:spect4j}
\end{eqnarray}
The last equality is a further assumption which implies an equation 
to be solved:
\begin{eqnarray}
-\frac{\frac{d\Delta(r)}{dr}}{E\!+\!M\!-\!\Delta(r)}
=-x\,\;\frac{(2l\!+\!5)\,\omega^2M}{E\!+\!M}\;\frac{r\,(\alpha^2\,r^2 -z)}{
\Big(\alpha^2\,r^2(2l\!+\!3\!+\!y)-(2l\!+\!1)\,y -\alpha^4\,r^4\Big)}\,.
\label{eq:spect4k}
\end{eqnarray}
The quantities $y$ and $z$ are two constants that have to be determined. 
The first one, which corresponds to a node in the large component 
of the wave function, is known at $x=0$ 
($y=\frac{2l+3}{2}$) and at $x=1$ ($y=\frac{2l+5}{2}$). The second one 
corresponds to a node in the small component  of the wave function
and is known at  $x=1$ where it has a simple expression ($z=\frac{2l+1}{2}$). 
It could be chosen to simplify part of the calculations.
Both $y$ and $z$ enter  Eq. (\ref{eq:GG}) to be solved. 
Looking at this equation, 
it is found that the coefficients of the terms proportional to $r^{l+5}$ 
and $r^{l-1}$ vanish with the choice we made for the wave function  
(using Eq. (\ref{eq:alpha}) in the first case 
and the low-$r$ behavior, $r^{l+1}$, in the second one).
The vanishing of coefficients of terms $r^{l+3}$ and $r^{l+1}$ 
supposes the conditions:
\begin{eqnarray}
&&E^2\!-\!M^2 - (2l\!+\! 7)\;\alpha^2+x \,\frac{(2l\!+\!5)\,\omega^2M}{E+M}=0 \,,
 \nonumber\\
&&(E^2\!-\!M^2)\,y-(2l\!+\!3)(2+y)\,\alpha^2+x \,\frac{(2l\!+\!5)\,\omega^2M}{E+M}
\;z=0\,.
\end{eqnarray}
The first of them can be cast in the general form given by 
Eq. (\ref{eq:specmn2}). The second one allows one 
to get $y$ once a choice has been made for $z$. This last quantity 
enters the determination of $\Delta(r)$ and, therefore, the potential 
which allows one to make the transition between $x=0$ and $x=1$, where 
it identifies to the one used in Chen {\it et al.}'s  work.
A particular choice consists in taking for  $z$ the value 
of $\alpha^2 r^2$ that cancels the denominator at the r.h.s. 
of Eq. (\ref{eq:spect4k}), the other value being denoted $z'$. 
The value of $y,\;z,\;z'$ so obtained are given by:
\begin{eqnarray}
y&=&\frac{1}{8\alpha^2(\Delta\! E^2\!+\!4\alpha^2)}
\Bigg((2l\!+\!3)\,((\Delta\! E^2)^2+6\Delta\! E^2\alpha^2+16\alpha^4)
-(2l\!+\!1)(\Delta\! E^2)^2
\nonumber\\
&& \hspace*{3cm}
-2\Delta\! E^2   \sqrt{(\Delta\! E^2\!+\!(2l\!+\! 7)\alpha^2)^2
+4(2l\!+\!1)(\Delta\! E^2\!+\!2\alpha^2)\alpha^2     } \Bigg)\, ,
\nonumber\\
z&=&\frac{1}{8\alpha^2}\Bigg((2l\!+\!3)(\Delta\! E^2+6\alpha^2)
-(2l\!+\!1)\Delta\! E^2
\nonumber\\
&& \hspace*{3cm}
-2   \sqrt{(\Delta\! E^2\!+\!(2l\!+\! 7)\alpha^2)^2
+4(2l\!+\!1)(\Delta\! E^2\!+\!2\alpha^2)\alpha^2     } \Bigg)\, ,
\nonumber\\
z'&=&2l+3+y-z
\nonumber\\
&=&\frac{1}{2(\Delta\! E^2\!+\!4\alpha^2)}
\Bigg((2l\!+\!3)(\Delta\! E^2+6\alpha^2)+(2l\!+\!1)\Delta\! E^2
\nonumber\\
&& \hspace*{3cm}
+2\sqrt{(\Delta\! E^2\!+\!(2l\!+\! 7)\alpha^2)^2
+4(2l\!+\!1)(\Delta\! E^2\!+\!2\alpha^2)\alpha^2     } \Bigg)\, ,
\end{eqnarray}
where $\Delta\! E^2= E^2\!-\!M^2-(2l\!+\! 7)\alpha^2=
-x \,\frac{(2l\!+\!5)\,\omega^2M}{E+M}$ ($\Delta\! E^2=0$ 
at $x=0$ and $\Delta \!E^2=-2$ at $x=1$).
The quantity   $\Delta(r)$ can then be calculated easily from the following 
equation: 
\begin{eqnarray}
-\frac{\frac{d\Delta(r)}{dr}}{E\!+\!M\!-\!\Delta(r)}
=-x\,\;\frac{\omega^2M}{E\!+\!M}\;\frac{(2l\!+\!5)\;r}{(z'\!-\!\alpha^2\,r^2)}
\label{eq:spect4l}
\end{eqnarray}
The solution takes a relatively simple form. It is similar 
to Eq. (\ref{eq:compae}) for the $n=1$ case and is given by: 
\begin{eqnarray}
E\!+\!M\!-\!\Delta(r)=\;(E\!+\!M)\;
\Big(1\!-\!\frac{\alpha^2r^2}{z'}\Big)^{ 
\frac{x (2l\!+\!5)\omega^2 M }{2\alpha^2 (E\!+\!M)}}\,.
\label{eq:spect4m}
\end{eqnarray}
It can be checked that the exponent is equal to 1 at $x=1$, 
so that the factor at the denominator in Eq. (\ref{eq:spect4j}) 
exactly cancels a similar factor $(z'\!-\!\alpha^2r^2)$ 
at the numerator (up to a constant factor). 

The solution for the small component is given by:
\begin{eqnarray}
F(r)=\frac{ (\frac{d}{dr}\!+\!\frac{\kappa}{r},\,G(r))}{E\!+\!M\!-\!\Delta(r)}
=  \frac{(2l\!+\!1)}{E\!+\!M}\;
\Big(1\!-\!\frac{\alpha^2r^2}{z'}\Big)^{(1- 
\frac{x (2l\!+\!5)\omega^2 M }{2\alpha^2 (E\!+\!M)})}
\;\frac{(1-\frac{\alpha^2r^2}{z})\,G(r)}{(1-\frac{\alpha^2r^2}{y})\,r} \,.
\label{eq:spect4}
\end{eqnarray}
%
\noindent
$\bullet$ {\it Case of states with $n=1$ 
and $j=l+1/2$: $1s_{1/2},\;1p_{3/2},\;1d_{5/2},\;1f_{7/2},\cdots$, 
($\tilde{l}=l\!+\!1$)}\\
The solution we retained in this case has been obtained from the observation
that there is some symmetry between Eq. (\ref{eq:GG}) for the large component 
and Eq.  (\ref{eq:FF}) for the small one. Most developments presented 
in the main text for the case  ($\tilde{l}=l\!+\!1$) can thus be transposed 
here. For the case  with $n=1$, we therefore look for a  small component 
with the following form
$F(r) \propto r^{(\tilde{l}+1)} \,{\rm exp}(-\alpha^2\,r^2/2)
\propto r^{(l+2)} \,{\rm exp}(-\alpha^2\,r^2/2)$, which is common to the
expected solutions at $x=0$ and $x=1$. 
The solution for the large component can be obtained  
using  its expression in term of the small one together with the expression 
of the transition potential given by Eq. (\ref{eq:compaed}): 
\begin{eqnarray}
G(r)=-\frac{ (\frac{d}{dr}\!-\!\frac{\kappa}{r},\,F(r))}{E\!-\!M\!-\!V(r)}
= -\frac{2\tilde{l}\!+\!1}{E\!-\!M}\;
\Big(1\!-\!\frac{\alpha^2\,r^2}{2\tilde{l}\!+\!1}\Big)^{(1- 
\frac{(1\!-\!x) (2\tilde{l}\!+\!1)\omega^2 M }{2\alpha^2 (E\!-\!M)})}\;
\;\frac{F(r)}{r}\, .
\label{eq:spect4i}
\end{eqnarray}
The above solution is consistent with the idea that the spin symmetry results from
the cancellation of $\Delta(r)$ while the pseudospin symmetry results from the
cancellation of $V(r)$, without any regard to the smallness of one component 
with respect to the other one. Solutions that are more in agreement with this
last observation have also been found. They nevertheless differ
from the solutions presented in the main text with many respects.
The equation allowing one to determine $E$ as a function of $x$ is more
complicated but the appearance of a node for the large component 
is more progressive. On the other hand, some modification of the normalization, 
that can be actually justified, is necessary to make it convergent.

\section{Transition from the SS to PSS limit in a relativistic model 
with finite potentials} \label{app:B}
In order to study the transition between two models A and B satisfying 
exact SS and PSS, respectively, we consider, in accordance 
with the definitions given in Subsect. \ref{ssec:rel1}, the potentials 
$V(r)=\Sigma_v(r)+\Sigma_s(r)$ and $\Delta(r)=\Sigma_v(r)-\Sigma_s(r)$ 
entering the Dirac equation (\ref{eq:dirac}):
  
\be
V(r)=(1-x)\frac{8M}{1+e^{-(r-r_0)/a}}\left[\frac{1}{1+e^{(r-r_0)/a}}
+\eta\frac{1}{1+e^{-(r-r_0)/a}}\right],
\label{V}
\ee

\be
\Delta(r)=x\frac{4M}{1+e^{-(r-r_0)/a}},
\label{D}
\ee
where, we take, somewhat arbitrarily,
$a=0.1$, $M=10$ fm$^{-1}$, $r_0=2$ fm, and
\be
\eta=\frac{4(x_0-x)}{1+100(x-x_0)^2},
\label{eta}
\ee
with (see below)
\be
x_0=\frac{E+M}{4M}.
\label{x0}
\ee

\noindent
Thus, for $x=0$ we recover the model A, which satisfies the exact spin symmetry, 
whereas for $x=1$, we have the model B, which satisfies the exact pseudospin
symmetry. 
We do not think our main conclusions will depend very much on our specific 
election of parameters. Like for Chen {\it et al.}'s model \cite{Chen02}, 
the strengths of the potential $V(r)$ at $x=0$ and $\Delta(r)$ at $x=1$ 
are small in the vicinity of  $r=0$. In the large-$r$ limit, they are 
however bounded by an upper finite value instead to increase infinitely.  
The value of $e=E-M$ is then necessarily positive at $x=0$ or $x=1$.  
In what follows, we may omit, by simplicity, the explicit dependence 
of $V(r)$ and $\Delta(r)$ on the space coordinate $r$.

To facilitate the understanding of the solutions of the Dirac equation
(\ref{eq:dirac}) with these potentials, it is useful to consider 
the corresponding equivalent Schr\"odinger equation for the large component
(see Ref. \cite{Mar08} for details). In this equation, with the potential 
$\Delta$ that becomes constant for $r\rightarrow\infty$,
the central potential ($U(r)\equiv U$) at $r\rightarrow\infty$ 
can be approximated as
\be
U(r\rightarrow\infty)=V+\frac{1}{2M}\left[e(V+\Delta)-e^2-V\Delta\right],
\label{U}
\ee
\noindent
where $e\equiv E-M$.

The last term in Eq. (\ref{V}), $\propto \eta$, makes $V(r\rightarrow\infty)$ 
different from zero. Its sign changes at $x=x_0$ to make the central potential 
$U(r)$ attractive at large values of $r$. The factor 4 entering the expression 
of $\eta$ controls the ``speed" of the transition from the SS to the PSS limit 
as $x$ varies. The factor $100$ (together with the factor $4$) determines 
the value of $V$ at $r\rightarrow\infty$ for a given value of $x$.

The potential $V(r\rightarrow\infty)$ can be written in terms of $U$ 
and $\Delta$ as, 
\be
V(r\rightarrow\infty)=\frac{2M(U/e)-(\Delta-e)}{2M-(\Delta-e)}e.
\label{V2}
\ee

\noindent
Notice that $x_0$ is the value of $x$ for which the denominator 
$2M-(\Delta-e)=0$. The potential $V$ is chosen so that it remains always finite 
at $r\rightarrow\infty$, in particular for $x=x_0$.
In this case, necessarily, $U(r\rightarrow \infty)=e$ for $x=x_0$. 
Thus, the system for $x\rightarrow x_0$ spreads all over the space. 
However, with the choice for the potentials $V$ and $\Delta$, 
it is possible to find bound states for $x$ very close to $x_0$ and, 
in most of cases, the SP energies vary smoothly with $x$ around $x_0$. 
Only when the value  of the SP-state energy is high enough and $x_0$ 
approaches the true value of $x$, for which the transition from the SS 
to the PSS limit is produced, in practice, the SP energy may vary quite 
sharply for states with $l$ large and $j=l-1/2$
(see the SP energies for the states $f_{5/2}$ and $g_{7/2}$ 
in Fig. \ref{fig_hwsrel}).


\end{document}